%% file: template.tex
\documentclass{article}

\usepackage{arxiv}
\usepackage{amsmath}

\usepackage[utf8]{inputenc} 
\usepackage[T1]{fontenc}    
\usepackage{hyperref}       
\usepackage{url}            
\usepackage{booktabs}       
\usepackage{amsfonts}       
\usepackage{nicefrac}       
\usepackage{microtype}      
\usepackage{cleveref}       
\usepackage{lipsum}         
\usepackage{graphicx}
\usepackage{natbib}
\usepackage{doi}
\usepackage{todonotes}
\usepackage{subcaption}
\usepackage{rotating}
\usepackage{booktabs}
\usepackage{tikz}
\usetikzlibrary{arrows.meta, positioning, fit}
\usepackage{rotating}
\usepackage{soul}
\usepackage[export]{adjustbox}

\tikzset{
block/.style={rectangle,
rounded corners=3pt,
minimum width=2.0cm,
minimum height=0.8cm,
align=center,
font=\footnotesize,
draw=black!40
},
data/.style={block, fill=blue!12},
process/.style={block, fill=gray!18},
conv/.style={block, fill=orange!25},
pool/.style={block, fill=purple!20},
output/.style={block, fill=red!20},
arrow/.style={thick, -{Latex[length=2.5mm]}, draw=black!70}
}

\title{Towards Interpretable Damage Detection based on Aerodynamic Pressure Measurements}

\newif\ifuniqueAffiliation
\uniqueAffiliationtrue

\ifuniqueAffiliation 
\author{ \href{https://orcid.org/0009-0003-7384-9371}{\includegraphics[scale=0.06]{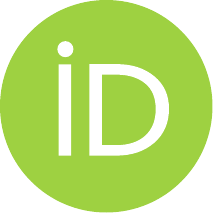}\hspace{1mm}Philip Franz} \\
	German Aerospace Center (DLR)\\
	Institute for the Protection of Terrestrial Infrastructures\\
	Rathausallee 12, 53757 St. Augustin, Germany \\
	\texttt{philip.franz@dlr.de} \\
	\And
	\href{https://orcid.org/0000-0002-2814-0027}{\includegraphics[scale=0.06]{orcid.pdf}\hspace{1mm}Max von Danwitz} \\
	German Aerospace Center (DLR)\\
	Institute for the Protection of Terrestrial Infrastructures\\
	Rathausallee 12, 53757 St. Augustin, Germany\\
	\texttt{max.vondanwitz@dlr.de} \\
    \And
	\href{https://orcid.org/0000-0002-0895-6766}{\includegraphics[scale=0.06]{orcid.pdf}\hspace{1mm}Gregory Duthé} \\
	Institute for Structural Engineering\\
	ETH Zürich\\
	Stefano-Franscini-Platz 5, 8093 Zürich, Switzerland\\
	\texttt{duthe@ibk.baug.ethz.ch} \\
    \And
	\href{https://orcid.org/0000-0002-8820-466X}{\includegraphics[scale=0.06]{orcid.pdf}\hspace{1mm}Alexander Popp} \\
	Institute for Mathematics and Computer-Based \\ Simulations (IMCS)\\
	University of the Bundeswehr Munich\\
	Werner-Heisenberg-Weg 39, 85577 Neubiberg, Germany\\
	\texttt{alexander.popp@unibw.de} \\
    \And
	\href{https://orcid.org/0000-0002-6870-240X}{\includegraphics[scale=0.06]{orcid.pdf}\hspace{1mm}Eleni Chatzi} \\
	Institute for Structural Engineering\\
	ETH Zürich\\
	Stefano-Franscini-Platz 5, 8093 Zürich, Switzerland\\
	\texttt{chatzi@ibk.baug.ethz.ch} \\
}
\else
\usepackage{authblk}

\newbox{\orcid}\sbox{\orcid}{\includegraphics[scale=0.06]{orcid.pdf}} 
\fi

\renewcommand{\shorttitle}{} 

\hypersetup{
pdftitle={Towards Interpretable Damage Detection Based On
Aerodynamic Pressure Measurements},
pdfauthor={Philip Franz, Max von Danwitz, Gregory Duthé, Alexander Popp, Eleni Chatzi}, 
pdfkeywords={First keyword, Second keyword, More},
}

\begin{document}
\maketitle

\begin{abstract}
	The increasing flexibility of modern large wind turbine blades necessitates cost-efficient and reliable structural monitoring solutions. For this purpose, we propose to use aerodynamic pressure measurements obtained via Aerosense, a novel, non-intrusive and economical sensing system. In former work ~\citep{Franz25}, we investigated the potential of aerodynamic pressure measurements for structural damage detection on elastic and aerodynamically loaded structures. An experimental campaign was conducted on a NACA 633418 airfoil mounted on a vertically vibrating cantilever beam within an open wind tunnel. Structural damage was introduced progressively through controlled saw cuts near the beam support. Aerodynamic pressure distributions were recorded under varying inflow conditions and structural states. Based on this data set, we developed a convolutional neural network to detect structural damage and classify its severity using only aerodynamic pressure signals. The results demonstrate that pressure measurements can effectively enable real-time detection and quantification of damage in elastic, beam-like structures subjected to mildly turbulent flow and varying operational conditions. Recognizing the limitations of pure black-box classification, in this study, we further incorporate physics-based insights and explainable machine learning methods to interpret how structural damage influences both the dynamic response and the aerodynamic pressure field. This leads to an enhanced damage detection pipeline, aiming to improve transparency, robustness, and physical consistency in data-driven monitoring of elastic, aerodynamically loaded structures.
\end{abstract}


\section{Introduction}
Deep neural networks, such as convolutional or transformer architectures, have achieved remarkable success in pattern recognition tasks for multivariate time series, like classification, prediction or anomaly detection~\citep{Fawaz.2019, Mohammadi24, Kong.2025, Zamanzadeh25}. 
This capability makes them particularly attractive for domains that rely on extracting information from dense, multi-sensor measurements, such as structural health monitoring (SHM)~\citep{Wang.2024, Cha.2024}.
However, relying on opaque black-box models for high-stakes decisions in SHM is risky, as deep neural networks typically lack interpretability. Furthermore, it remains unclear whether these models generalize reliably to real-world environments where operating conditions differ from the training distribution. Compounding this, inferring states from sparse and noisy measurements can be ill-posed, admitting a distribution of plausible solutions rather than a unique answer~\citep{vadeboncoeur2026gabi}.
To address the lack of generalization and interpretability, the field of scientific machine learning~(SciML)~\citep{Baker.2019, Quarteroni.2025} or physics-informed machine learning has emerged~\citep{Karniadakis.2021}. SciML integrates physical laws or domain-specific knowledge directly into the learning scheme, aiming to ensure adherence to the underlying physics, generalization beyond the training domain and to enhance the interpretability of the learned representations~\citep{Karniadakis.2021, Cicirello24}.
However, SciML requires robust approaches and detailed physical insights and governing equations, which may not always be available for real-world systems. In such cases, explainable artificial intelligence~(XAI) offers an alternative for increasing the transparency of black-box models.
In our recent study~\citep{Franz25}, we suggested a convolutional neural network (CNN)-based approach to detect and rank structural damage based on aerodynamic pressure measurements (APM) over the section of an airfoil, acquired with the Aerosense system~\citep{Barber22}. 
Despite its accuracy, the CNN's black-box nature limits its practical utility and prevents a thorough understanding of the underlying mechanism.
Therefore, we employ an XAI method to enhance the interpretability of our approach and to identify damage-sensitive features within the APM. XAI methods can generally be categorized as ante-hoc or post-hoc, global or local, and model-specific or model-agnostic~\citep{Theissler22}. 
For this study, we select Integrated Gradients~(IG)~\citep{Sundararajan17}, a model-agnostic, post-hoc attribution method~\citep{Theissler22}, which is suited for multivariate time series and provides attributions across all channels. While IG is a local method providing only sample-specific explanations, subsequence or instance-based approaches such as XCM~\citep{Fauvel.2021}, TSProto~\citep{Bobek.2025}, or ProtoTSNet~\citep{Makus.2025}, can offer global, data set-wide explanations via class-specific subsequences or prototypes. However, these methods require training new models, whereas our objective is to interpret the existing CNNs presented in~\citep{Franz25}.
The paper is organized as follows: Section~\ref{sec:physics-insights} details the experimental setup, the previously trained CNNs and effects of damage on the aerodynamic pressure distribution (APD). Section~\ref{sec:cnn_attribution} presents the IG method and baselines, analyzing the resulting attribution maps. Section~\ref{sec:class} describes the evaluation and retraining of the CNNs based on insights from the attribution analysis, while Section~\ref{sec:disc} provides a physics-grounded discussion of the results. We conclude with a summary of the main findings in Section~\ref{sec:concl}.
%
\section{Experimental Setup \& Effects Of Damage}\label{sec:physics-insights}
%

\subsection{Experimental setup} \label{subsec:setup}
\begin{figure}[h]
    \centering
	\includegraphics[width=\linewidth]{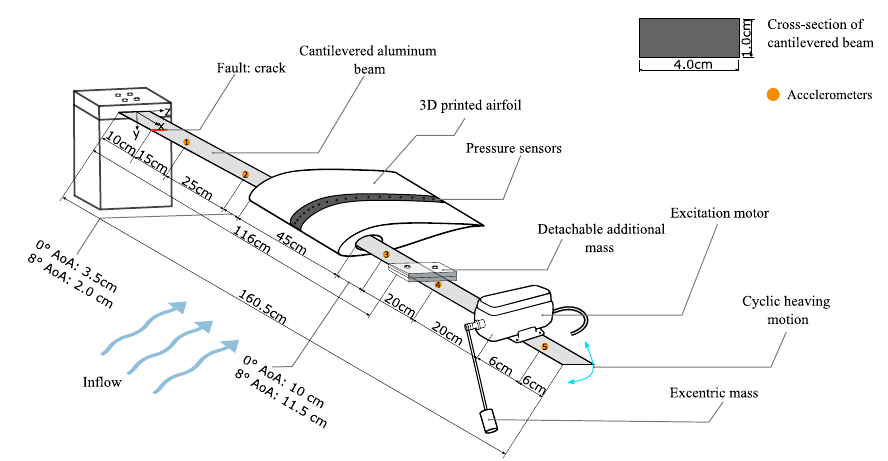}
	\caption{The experimental setup features an airfoil on a cantilever beam with an excitation motor and removable mass, placed in front of the outlet of the wind tunnel. The Aerosense sensor node records aerodynamic pressure on the airfoil, and accelerometers measure beam vibration. Beam dimensions are indicated in the upper right. Figure taken from~\citep{Franz25}.}
	\label{fig:setup}	
\end{figure}

\begin{figure}[h]
    \centering
	\includegraphics[width=0.7\linewidth]{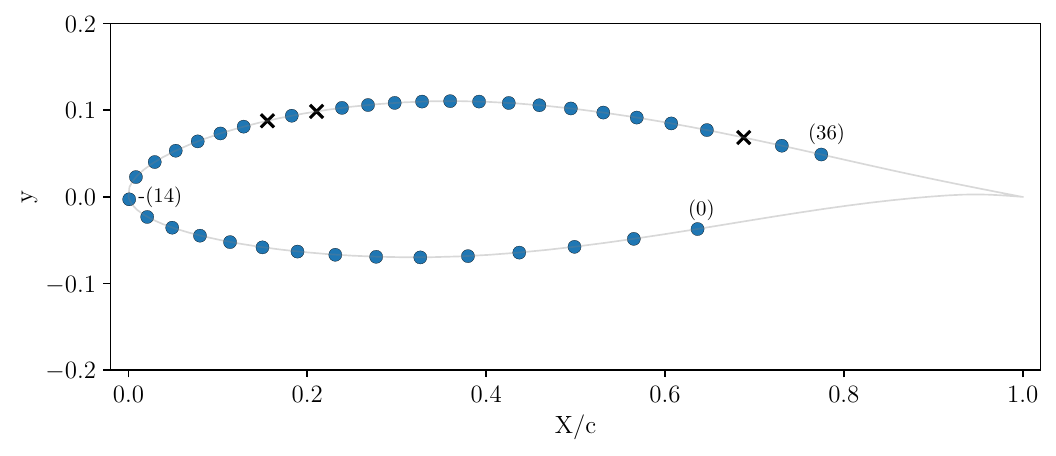}
	\caption{Distribution of barometers on the airfoil surface. The blue, circular markers indicate working sensors; the black crosses indicate malfunctioning sensors. The numbers of the sensors at the trailing edge and at the leading edge are indicated by the numbers in the brackets.}
	\label{fig:sensors_airfoil}	
\end{figure}

The experimental setup (Figure~\ref{fig:setup}) consists of a NACA~633418 airfoil mounted on a flexible aluminum cantilever beam within the open test section of the EPFL Unsteady Flow Diagnostics Laboratory wind tunnel. The wind tunnel test section measures 40~cm$\times$40~cm and operates up to 35~m/s, while the beam itself features a rectangular cross-section (1~cm~×~4~cm). Artificial damage is introduced by transversely sawing the beam in z-direction near the support to reduce local stiffness; the severity is quantified by the cut length across five levels (0\%, 12.5\%, 25\%, 37.5\%, and 50\% of the beam width). An additional damage scenario is considered by attaching a mass to the undamaged beam between the airfoil and the motor (see Figure~\ref{fig:setup}). Periodic excitation is applied at the beam's tip via a motor-driven eccentric mass, primarily inducing flap-wise bending, although torsional motion also occurs. The Aerosense system~\citep{Barber22} records the aerodynamic pressure distribution at 40 locations on the airfoil surface (only 37 were functioning correctly, see Figure~\ref{fig:sensors_airfoil}) at the airfoil's mid-span with a sampling rate of 100 Hz. Ultimately, this cantilever setup serves as a proxy for a rotating wind turbine blade oscillating flap-wise due to gravitational loading.
\noindent Table~\ref{tab:boundary_conditions} summarizes the eight test series (TS) generated by combining the two angles of attack (AoAs), wind speeds~$V$, and heaving frequencies~$f_h$. Each TS comprises six structural states resp. damage classes: five crack lengths to simulate stiffness reduction and one added mass state (see Table \ref{tab:damage_classes}). We conduct three~150~s runs per configuration: the first~15~s feature aerodynamic loading only, while the remaining~135~s involve simultaneous loading by aerodynamic and periodic forces. More details on the experimental setup, the data acquisition system and data normalization, can be found in Section~2 of~\citep{Franz25}.

\begin{table}[b]
\hspace{-8pt}
\centering
\caption{Boundary conditions of the eight test series.}
\begin{tabular}{@{}llcccccccc@{}}
\toprule
Test series            &  & 1   & 2   & 3   & 4   & 5   & 6   & 7   & 8   \\ \midrule
AoA [$^\circ$]       &  & 0   & 0   & 0   & 0   & 8   & 8   & 8   & 8   \\
approximate $f_h$ [Hz]        &  & 1.0 & 1.0 & 1.9 & 1.9 & 1.0 & 1.0 & 1.9 & 1.9 \\
v [m/s] &  & 12  & 24  & 12  & 24  & 12  & 24  & 12  & 24  \\ \bottomrule
\end{tabular}
\label{tab:boundary_conditions}
\end{table}

\subsection{Previous results with black-box approach} \label{subsec:cnn}
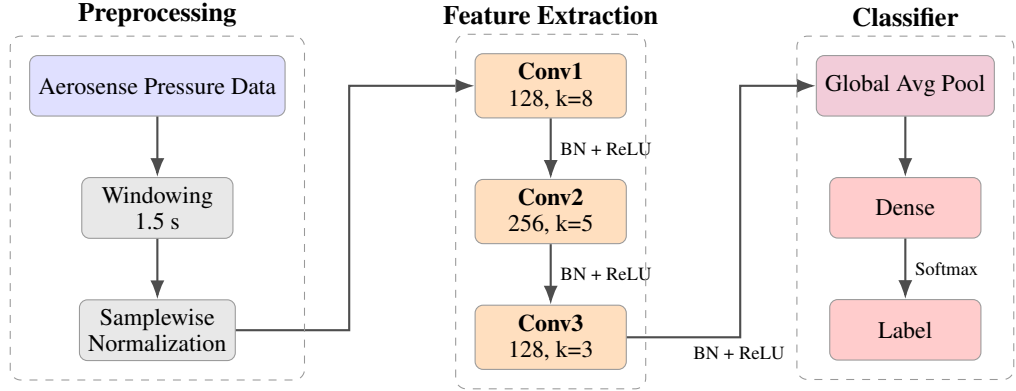
\begin{figure}[h]
\centering
\begin{tikzpicture}[node distance=0.8cm and 1.6cm]

\node[data] (raw) {Aerosense Pressure Data};
\node[process, below=of raw] (window) {Windowing \\ 1.5 s};
\node[process, below=of window] (norm) {Samplewise \\ Normalization};

\node[conv, right=2.5 of raw] (conv1) {\textbf{Conv1}\\128, k=8};
\node[conv, below=of conv1] (conv2) {\textbf{Conv2}\\256, k=5};
\node[conv, below=of conv2] (conv3) {\textbf{Conv3}\\128, k=3};
\node[pool, right=2.5 of conv1] (gap) {Global Avg Pool};
\node[output, below=of gap] (softmax) {Dense};
\node[output, below=of softmax] (label) {Label};

\draw[arrow] (raw) -- (window);
\draw[arrow] (window) -- (norm);
\draw[arrow] (norm.east) -- ++(1.5,0) |- (conv1.west);

\draw[arrow] (conv1) -- node[right,font=\scriptsize]{BN + ReLU} (conv2);
\draw[arrow] (conv2) -- node[right,font=\scriptsize]{BN + ReLU} (conv3);
\draw[arrow] (conv3.east) -- node[below right, font=\scriptsize]{BN + ReLU} ++(1.5,0) |- (gap.west);

\draw[arrow] (gap) -- (softmax);
\draw[arrow] (softmax) -- node[right,font=\scriptsize]{Softmax}(label);

\node[draw=black!40, dashed, rounded corners, fit=(raw)(norm), inner sep=0.25cm, label=above:{\textbf{Preprocessing}}
](prepro) {};

\node[draw=black!40, dashed, rounded corners, fit=(conv1)(conv3), inner sep=0.25cm, label=above:{\textbf{Feature Extraction}}](feat_ex){};

\node[draw=black!40, dashed, rounded corners, fit=(gap)(label), inner sep=0.25cm, label=above:{\textbf{Classifier}}]{};
\end{tikzpicture}

\caption{Preprocessing pipeline and CNN architecture for pressure signal classification.}
\label{fig:algo}
\end{figure}

The damage detection and severity rating algorithm proposed in~\citep{Franz25}, follows the CNN architecture (see Figure~\ref{fig:algo}) suggested by~\cite{Wang.2017}, comprising three convolutional layers with batch normalization, a global average pooling layer, and a fully connected layer. To prepare the data, the first~40~s and last~10~s of each measurement are discarded to mitigate transient effects. From the remaining~100~s,~89 overlapping~1.5~s windows are extracted to capture approximately three vibration cycles per sample and minimize the influence of signal trends. Finally, each sample is z-score normalized and split into training, validation, and test sets, resulting in the algorithm shown in Figure~\ref{fig:algo}.
In~\citep{Franz25}, we separated the data for the classification by AoA, as the pressure distribution depends strongly on this variable. With six structural states across two frequencies and two wind speeds (see Table~\ref{tab:boundary_conditions}), each AoA data set comprises~72 measurements. We employ a non-random split strategy: two of the three measurements for each boundary condition are consistently assigned to training, while the third is reserved for testing to ensure evaluation on unseen data. This
results in~48 experiments for training and~24 for testing. With the segmentation approach described above, this yields~4,272 training samples and~2,136 test samples. The training set is further divided into a~25\% validation set~(1068 samples) and a~75\%  training set~(3204 samples). An overview of the classification results on the test set of each split is given in Table \ref{tab:all_splits}. The classification accuracy per split is the average accuracy over all balanced classes. Further details on the CNN approach, the split of the data set and the classification results are available in Section 5~of~\citep{Franz25}. 

\begin{table}[t]
\centering
\hspace{-8pt}
\caption{Test set accuracy for all splits for~0°~AoA and~8°~AoA. The classification accuracy given for a single split is the average of the classification accuracy over the six balanced classes of that split.}
\begin{tabular}{ccccc}
\toprule
 AoA     &  Split 1 & Split 2 & Split 3 & Average\\ \midrule
 0° &  82.6\,\%  & 98.5\,\%  &  93.7\,\% & 91.6\,\%\\
 8° &  81.8\,\% &  96.0\,\%  &  89.8\,\% & 89.2\,\% \\ \bottomrule
\end{tabular}
\label{tab:all_splits}
\end{table}

\subsection{Effects of damage on  structural behavior and aerodynamic pressure field}\label{subsec:hyp}

\begin{figure}[h]
    \centering
	\def\svgwidth{0.8\linewidth}
    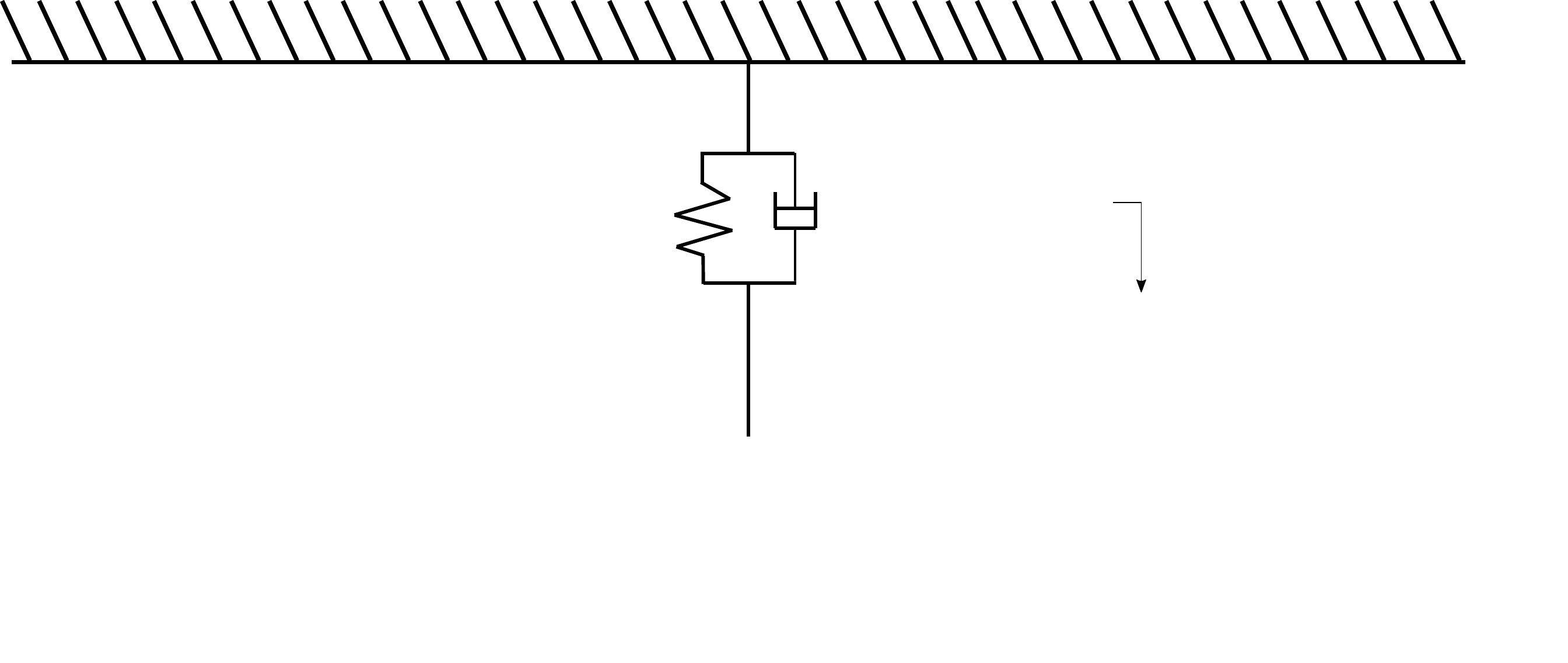
	\caption{Schematic blade section with a flap-wise degree of freedom $y$ and rotational degree of freedom $\phi$. M denotes the system's mass, $K_y$ and $D_y$ denote stiffness and damping in flap-wise direction and $K_\phi$ and $D_\phi$ denote the rotational stiffness and damping. V describes the inflow velocity, V$_{rel}$ describes the relative inflow velocity and $\alpha$ the AoA . The airfoil is subjected to lift and drag forces, denoted by $F_L$ and $F_D$.}
	\label{fig:airfoil_2dof}	
\end{figure}
\begin{figure}[h]
    \centering
	\includegraphics[width=0.85\linewidth]{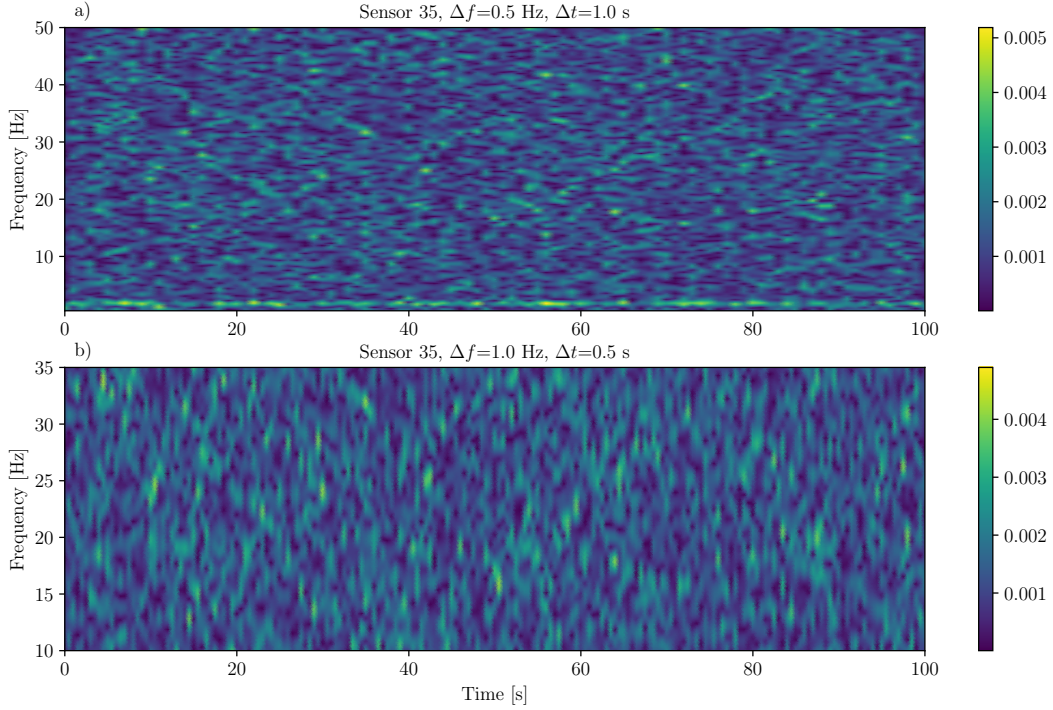}
	\caption{STFT spectra of sensor 35 (trailing edge) for TS 8 with a 50\% beam-width crack. Panels~a) and~b) show different time-frequency resolutions for the ranges 0.5-50~Hz and 10-35~Hz, respectively. No prominent energy bands or regular pulses are detected at~15~Hz or~30~Hz.}
	\label{fig:stft}	
\end{figure}

The airfoil section is approximated as a spring-mass-damper system (Figure \ref{fig:airfoil_2dof}) with a vertical~$y$ and a rotational~$\phi$ degree of freedom. The system is characterized by mass~$M$, stiffness and damping coefficients~$K_y$,$D_y$,$K_{\phi}$ and $D_{\phi}$, inflow velocities $V$,$V_{rel}$, AoA $\alpha$, and the aerodynamic lift and drag force, $F_L$ and $F_D$. Stepwise cutting of the beam in the z-direction introduces artificial damage, influencing the dynamics as follows:
\noindent First, the equilibrium position shifts downward. The cut reduces the local cross-sectional area and the flexural moment of inertia, resulting in a downward y-displacement of the cantilever beam that increases along the longitudinal axis.
Additionally, the cut also alters the center of shear. Due to the eccentricity of the tip mass and airfoil, the beam exhibits a small initial twist; this torsional rotation~$\phi$ increases with the cut length as the torsional moment of inertia is reduced at the site of damage.
\noindent Secondly, as the crack length increases and the loading is not altered, the amplitudes of both flexural and torsional vibrations increase. Furthermore, these oscillations occur around an equilibrium position that becomes progressively more deflected and twisted. As discussed in~\cite{Franz25}, the torsional vibrations are expected to be significantly smaller than the vertical vibrations; however, both magnitudes increase as the damage progresses. 
\noindent Regarding the impact of damage on the APD, the increased twist of the equilibrium state and larger torsional and vertical vibration amplitudes are expected to significantly influence the pressure distribution, whereas the increased inclination of the airfoil in the y-direction is unlikely to have a substantial effect. Specifically, airfoil twist and torsional vibrations alter the AoA. The twist induces a static offset relative to the undamaged state, while torsional vibrations increase AoA fluctuations. Additionally, vertical oscillations induce a time-varying velocity component~$\dot{y}$ of the airfoil, resulting in corresponding variations in the AoA and APD. 
Moreover, torsional and flexural vibrations may enhance vortex shedding at the airfoil's trailing edge, further altering the APD. To estimate the vortex shedding frequency $f_{vs}$, we employ the Strouhal number $St = \frac{f_{vs}D}{V}$~\citep{ModarresSadeghi.2021}. Given the Reynolds regime of our experiments (see Section 3.3 of~\citep{Franz25}), we assume $St\approx0.2$ following~\cite{ModarresSadeghi.2021}, although an airfoil is not truly a bluff body. Using the airfoil chord length for $D$ following~\cite{Chang.2022}, the estimated shedding frequencies are $f_{vs} \approx \frac{St\times V}{D} = \frac{0.2 \times 12 \frac{m}{s}}{0.16 m} = 15$ Hz for $V=12$ m/s and $f_{vs}\approx30$ Hz for $V=24$ m/s.
Analysis of signal spectra from TS~8, the highest loading case, with a~50\% crack length revealed no significant energy bands or regular pulses at the predicted~15~Hz or~30~Hz frequencies (Figure~\ref{fig:stft}). The spectra were computed via SciPy's Short-Time Fourier Transform~(STFT) function~\citep{SciPy} for sensor~35, using two resolutions; panel~a) covers~0.5-50~Hz ($\Delta t=1.0$ s, $\Delta f=0.5$ Hz), while panel~b) focuses on~10-35~Hz ($\Delta t=0.5$ s, $\Delta f=1.0$ Hz) to better detect short-duration shedding events. Aside from the already known band at~$\approx1.9$~Hz, no consistent pulses were observed. Since TS~8 represented the most extreme conditions of the campaign, we conclude that significant vortex shedding did not occur in any of the experiments.
Finally, a second type of vortex shedding, induced by the oscillations of the airfoil, may occur and intensify as structural damage increases. Since this phenomenon is governed by the motion of the airfoil, its frequency is expected to coincide with the vibration frequency. Thus, however, this second type of vortex shedding is difficult to isolate within the frequency spectra.
%
\section{CNN Analysis Based On Integrated Gradients}\label{sec:cnn_attribution}
%
\subsection{Integrated gradients}\label{subsec_ig_int}
\begin{figure}[h]
    \centering
	\includegraphics[width=0.9\linewidth]{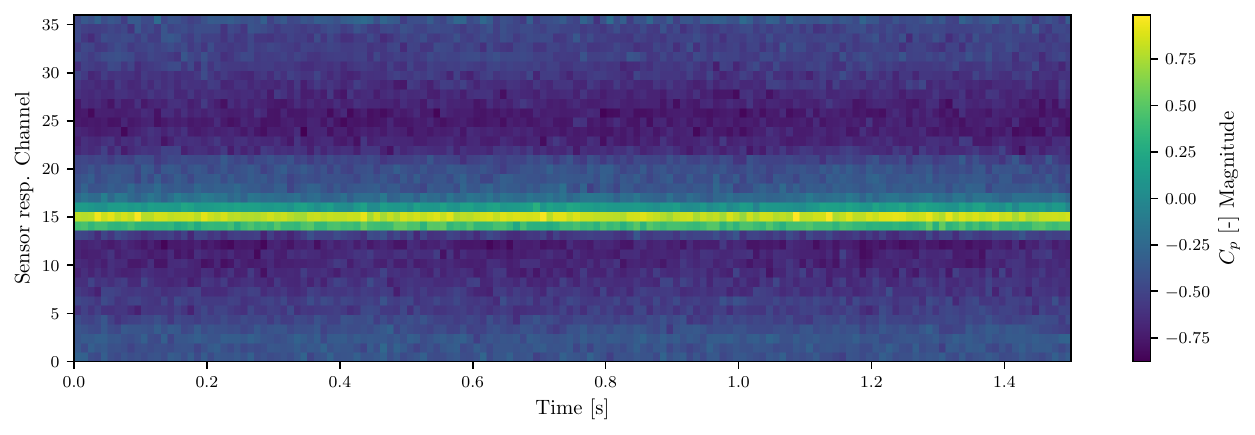}
	\caption{Exemplary multivariate time series sample (sample 14) $\in R^{37 \times 150}$ of the correctly classified samples of the validation set of split 1, 0° AoA.}
	\label{fig:signal}	
\end{figure}

\begin{figure}[h]
    \centering    
	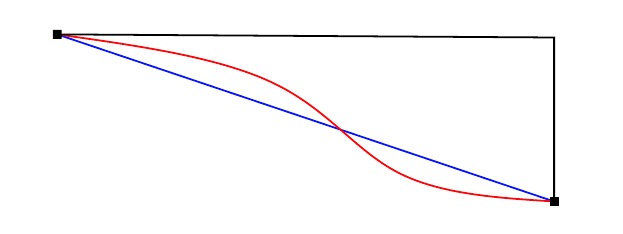
	\caption{Example of possible integration paths in two dimensions between the baseline $x'$ and the full input sample $x$. IG takes the linear integration path $P_2$ (blue). Image inspired by~\cite{Sundararajan17}.}
	\label{fig:paths}	
\end{figure}

%
To mitigate the opacity of the CNN previously described and to identify which input features drive classification into a specific damage class, we subsequently perform an IG analysis. IG is an attribution method for neural networks introduced in~\citep{Sundararajan17}. We emphasize, however, that IG highlights sensitivity of the model output to input perturbations, rather than providing a causal or physically unique explanation.
For a multivariate time series sample $x\in\mathbb{R}^{C\times T}$ (compare Figure \ref{fig:signal}), where $C=37$ denotes the number of channels and $T=150$ the number of time steps, IG evaluates the importance of an element $x_{c,t}$, with $c\in[1,C]$ (mapped to the noncontiguous set of working sensors 0 to 36, see Figure \ref{fig:sensors_airfoil}) and $t\in[1,T]$, relative to a baseline $x'\in \mathbb{R}^{C \times T}$, for the classification function $F(x):\mathbb{R}^{C\times T}\rightarrow[0;1]^m$ encoded by the fully trained CNN. As the data set comprises six classes, $m=6$.
For this purpose, $x'$ is defined such that it excludes a specific signal property, e.g. the mean value of $x$, by modifying all elements $x'_{c,t}$ accordingly.
Then a linear path in the sample space $\mathbb{R}^{C\times T}$ (see Figure \ref{fig:paths}) 
is defined between the baseline $x'$ and the true sample $x$. 
Following Equation \ref{eq:ig}, the IG for element $x_{c,t}$ is then computed by determining the gradient of $F$ for each point on this path $x' + \gamma(x-x')$ with $\gamma \in [0;1]$, with respect to the $c,t$-th dimension and aggregating these:
\begin{equation}
   IG_{c,t}(x):= (x_{c,t} - x_{c,t}')\cdot\int_{\gamma=0}^{1} \frac{\partial F(x'+\gamma(x-x'))}{\partial x_{c,t}}d\gamma, \label{eq:ig}
\end{equation}
\noindent Computing $IG_{c,t}$ for all elements in $x$ yields the attribution map $IG(x) \in \mathbb{R}^{C \times T}$ for sample $x$ and baseline $x'$. The actual computation of $IG(x)$ is conducted via an approximation based on Riemann integration, where the integral is replaced by a sum over adequately sized intervals alongside the path between baseline and input sample. The number of steps $L$ along this path must be chosen large enough to allow for good approximation of the original integral. Sundararajan et al. recommend choosing $20<L<300$~\citep{Sundararajan17}. 
%
\subsection{Baselines and attribution analysis setup} \label{subsec_ig_bl}

\begin{figure}[h]
    \begin{center}
        \includegraphics[ width=\linewidth]{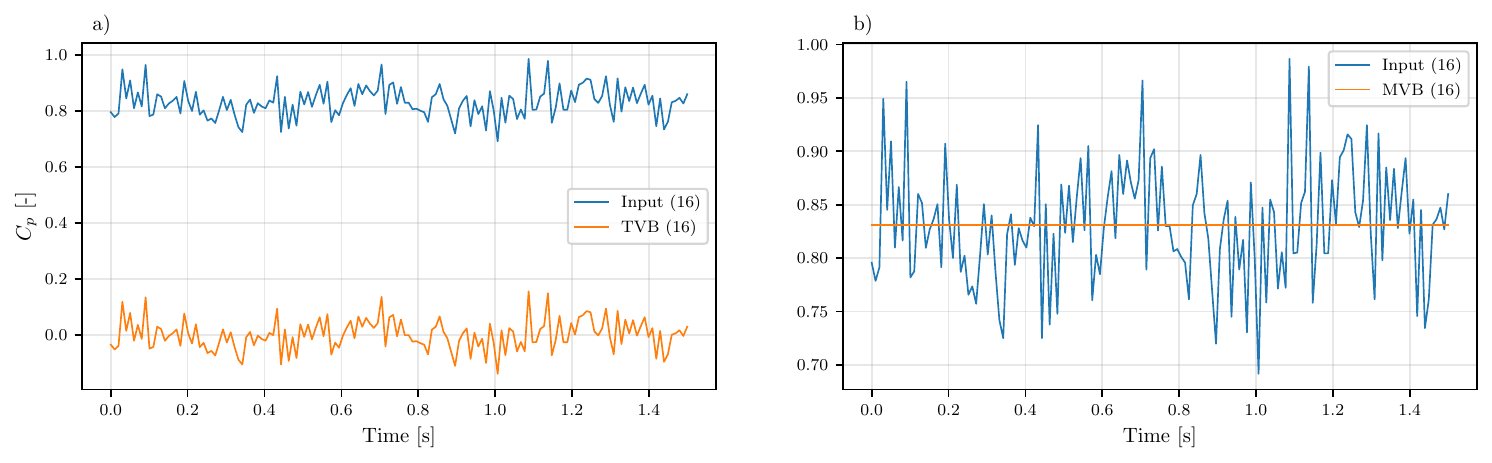}
        \caption{Visualizations of the TVB a) and MVB b) baselines for an exemplary sample and the signal of sensor~16. The APB, with the constant zero baseline is not displayed, as it consists of a constant line at zero for the duration of the sample.}
    \label{fig:baselines}
    \end{center}
\end{figure}

For our analysis, we implement IG with TensorFlow\footnote{Version 2.14.0 with Python 3.10.19} and choose the following baselines for our multivariate time series data: 

\begin{enumerate}
    \item Ambient Pressure Baseline (APB): The first baseline we employ consists of constant zeros across the entire sample. We adopt this baseline of an aerodynamic pressure coefficient $C_p=0$ to emulate a scenario with no aerodynamic loading or vibration of the airfoil, thus only accounting for atmospheric pressure and neglecting sensor noise. 
    
    \item Temporal Variations Baseline (TVB): In this second baseline, we remove the offset of the signal in each channel, thereby eliminating the relative magnitude relationships between all channels while preserving only the temporal variations (see panel~a) of Figure \ref{fig:baselines}).  
    
    \item Channel-wise Mean Value Baseline (MVB): In this third baseline, the time-varying signal in each channel is replaced by its mean value. As a result, the sample becomes temporally constant across all channels (see Panel~b) of Figure \ref{fig:baselines}). This baseline eliminates the individual dynamics of each channel while preserving the relative magnitude relationships among them.
    
\end{enumerate}

\noindent Subsequently, we compute attribution maps for correctly classified samples of the validation set to gain insight into the decision mechanism of the model. We restrict the attribution analysis to the validation sets of the two splits that yielded the lowest test set classification accuracy: split~1 for the~0°~AoA data set and split~1 for the~8°~AoA data set (see Table~\ref{tab:all_splits}).
For the~0°~AoA validation set~1067 of~1068 samples were correctly classified and for~8°~AoA validation set~1057 samples. 
As the signal windows are obtained by sliding a fixed‑length window over the raw aerodynamic pressure signal, they are not phase‑aligned. Consequently, attribution maps cannot be easily aggregated across all samples. The analysis that follows is therefore primarily qualitative and based on a small set of randomly chosen samples from each TS (see Table~\ref{tab:boundary_conditions}) and damage class rather than on the entire pool of correctly classified samples.
For each sample in the validation set, we compute the corresponding attribution values with~200~steps along the linear path in the sample space. The resulting attribution maps contain positive and negative gradients, where positive gradients push the model towards the predicted class, negative gradients push against the predicted class. 
In addition, we compute the sum~$\boldsymbol{c} \in \mathbb{R}^{37}$ of attributions per channel over the sample length (Equation~\ref{eq:att_vec}), to evaluate which channels contribute most to the classification of the sample.
\begin{equation}
    \boldsymbol{c} = \sum_{k=1}^{150} IG_{k}(x), \\ \label{eq:att_vec}
\end{equation}
\noindent where $IG_{k}(x) \in \mathbb{R}^{37}$ describes the vector of attributions over all channels at the timestep~$k$.

\subsection{Attribution maps for the ambient pressure baseline}
\begin{figure}[h]
    \centering
	\includegraphics[width=\linewidth]{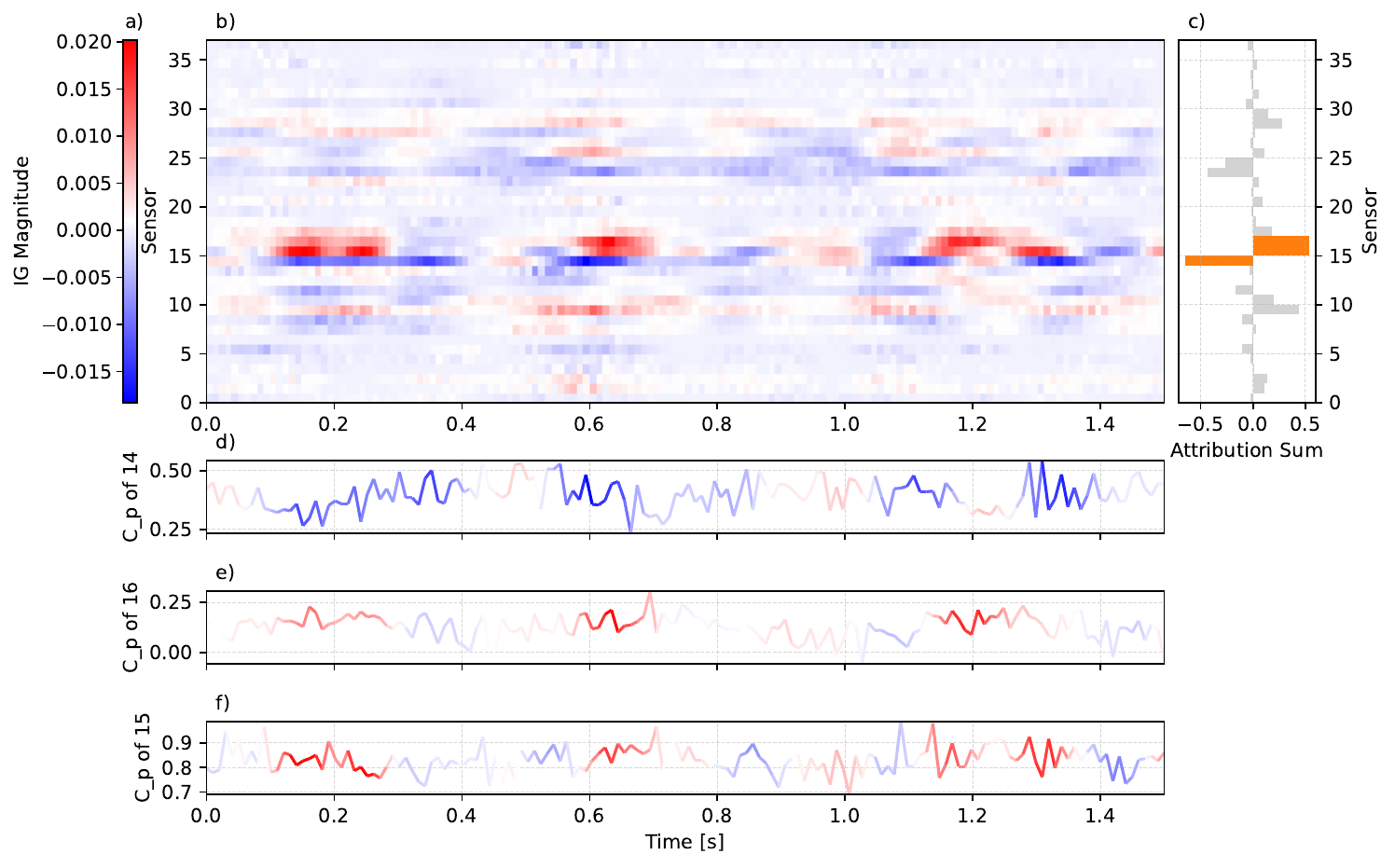}
	\caption{b) Attribution map for sample 14 (TS 1, damage class 0, split 1) using the APB. c) Sum of attributions per channel with the top three highlighted in orange. d)-f) Signals for the three highest‑attribution channels, colored by attribution score. The colorbar in a) applies to b) and d)-f).}
	\label{fig:sample_att_bs_apd}	
\end{figure}
\begin{figure}[h]
    \centering
	\includegraphics[width=\linewidth]{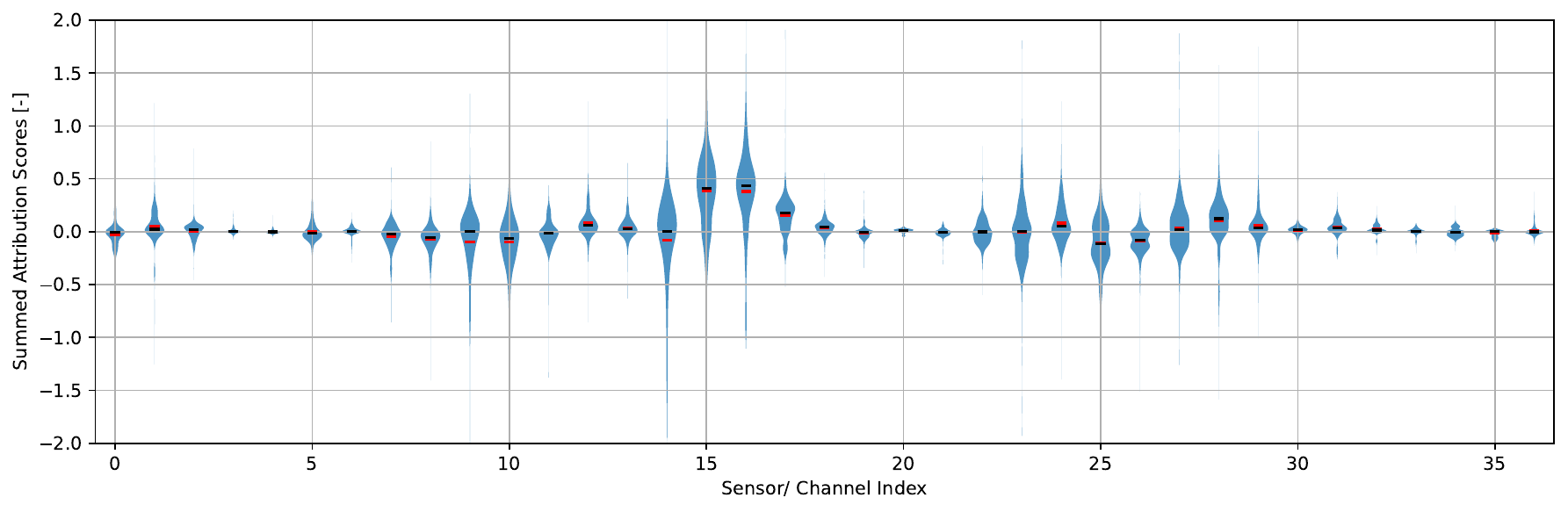}
	\caption{Violin plot presenting the distribution of the summed attribution vectors $c$ (see Equation \ref{eq:att_vec}) for all samples of the correctly classified validation set (split 1, 0° AoA) using the APB. The red bar in each violin marks the mean value; the black bar the median.}
	\label{fig:violin_apb}	
\end{figure}
\begin{figure}[h]
    \centering
	\includegraphics[width=\linewidth]{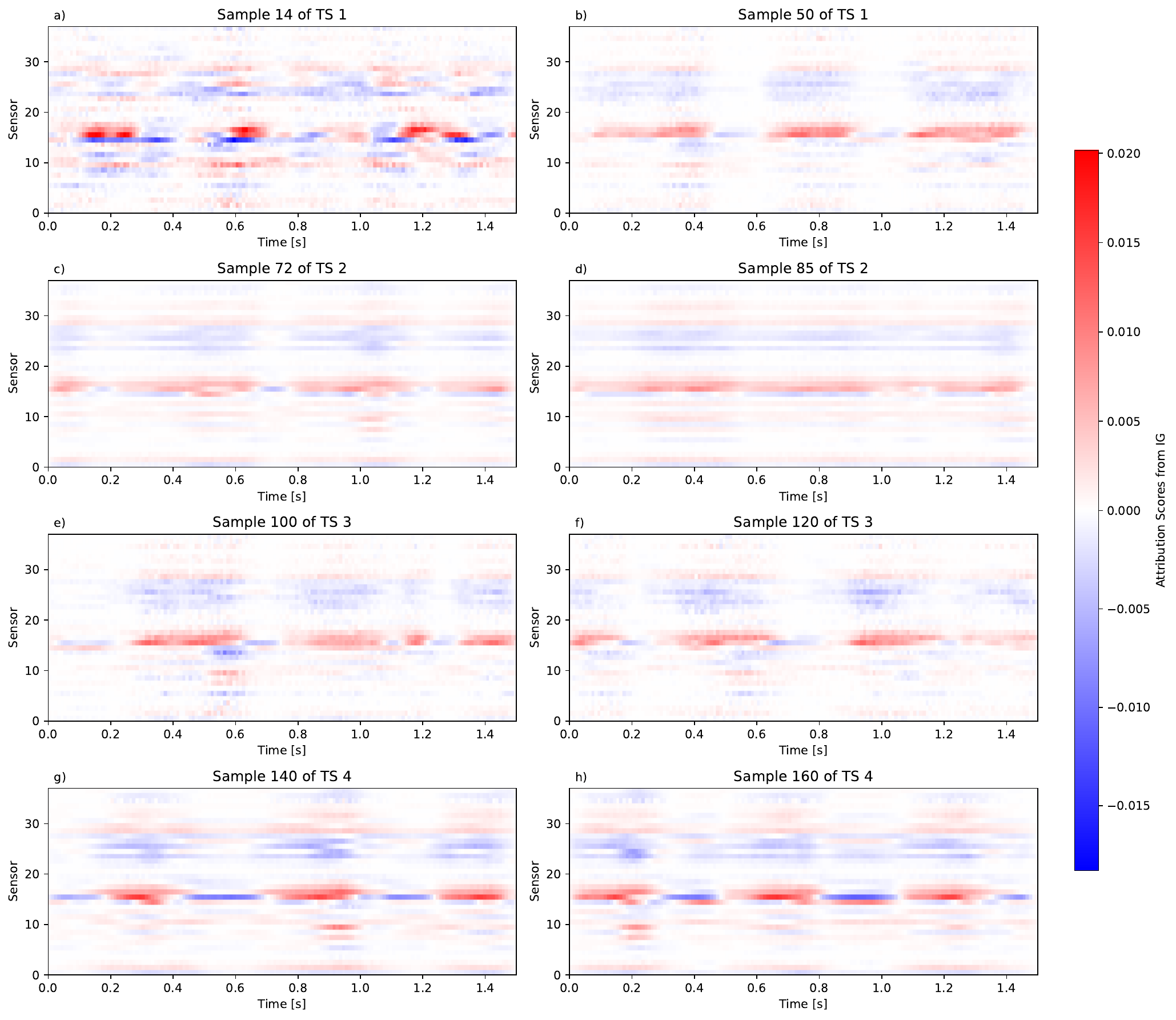}
	\caption{Overview of attribution maps based on the APB for different inflow conditions for damage class 0 and 0° AoA. Panels a) and b) belong to TS 1 (see Table \ref{tab:boundary_conditions}), c) and d) to TS 2, e) and f) to TS 3 and g) and h) to TS 4.}
	\label{fig:overview_ts_apb}	
\end{figure}
\begin{figure}[h]
    \centering
	\includegraphics[width=\linewidth]{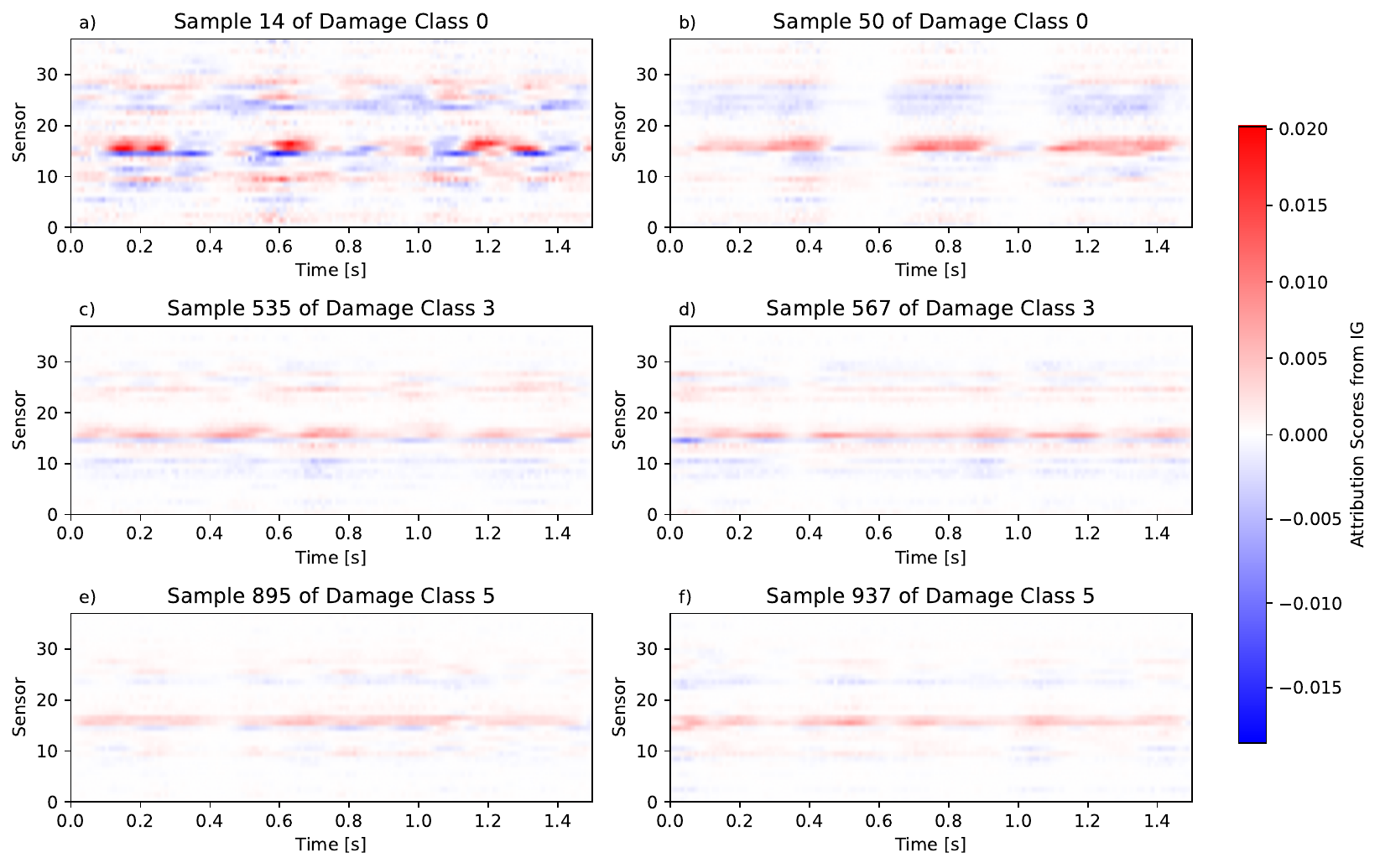}
	\caption{Overview of attribution maps based on the APB for different structural states for TS 1 and 0° AoA. Panels a) and b) belong to damage class 0 (no damage), c) and d) to damage class 3 (crack length of 25\% of the beam width) and e) and f) to damage class 5 (crack length of 50\% of beam width).}
	\label{fig:overview_dam_apb}	
\end{figure}
\begin{figure}[h]
    \centering
	\includegraphics[width=\linewidth]{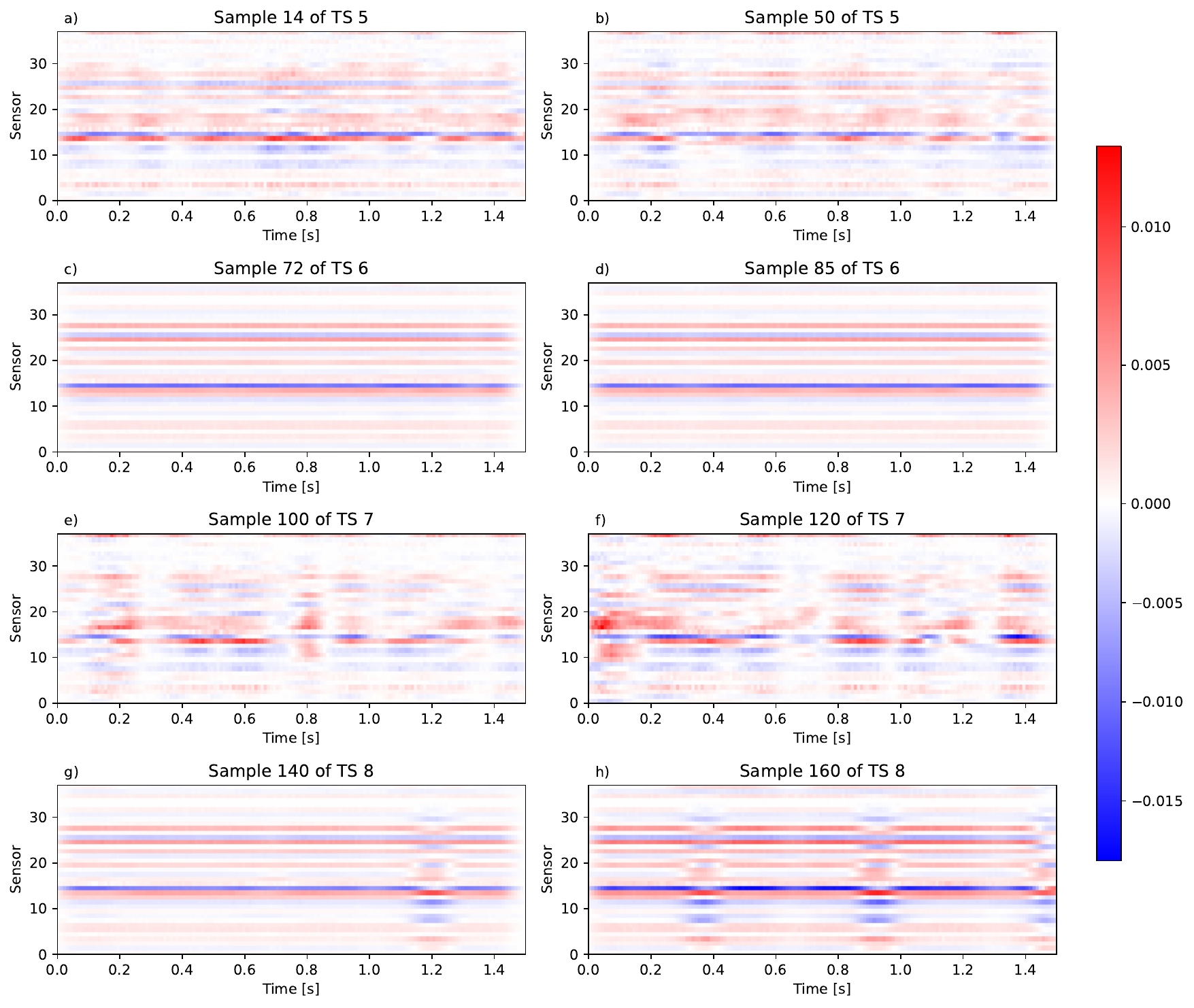}
	\caption{Overview of attribution maps based on the APB for different inflow conditions for damage class 0 and 8° AoA. Panels a) and b) belong to TS 5 (see Table \ref{tab:boundary_conditions}), c) and d) to TS 6, e) and f) to TS 7 and g) and h) to TS 8.}
	\label{fig:overview_ts_apb_8}	
\end{figure}
\begin{figure}[h]
    \centering
	\includegraphics[width=0.9\linewidth]{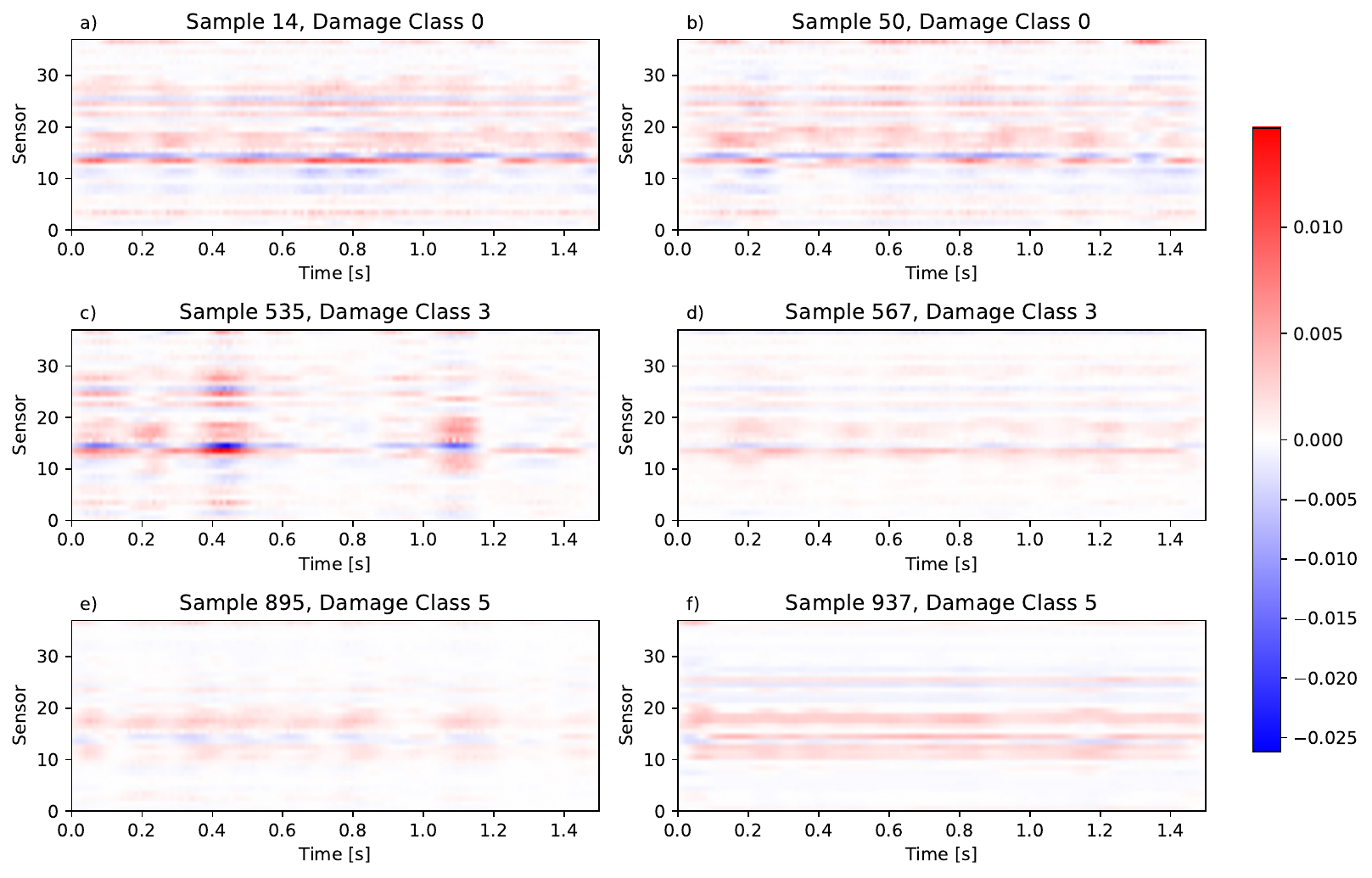}
	\caption{Overview of attribution maps based on the APB for different structural states for TS 5 and 8° AoA. Panels a) and b) belong to damage class 0 (no damage), c) and d) to damage class 3 (crack length of 25\% of the beam width) and e) and f) to damage class 5 (crack length of 50\% of beam width).}
	\label{fig:overview_dam_apb_8}	
\end{figure}
\begin{figure}[h]
    \centering
	\includegraphics[width=0.88\linewidth]{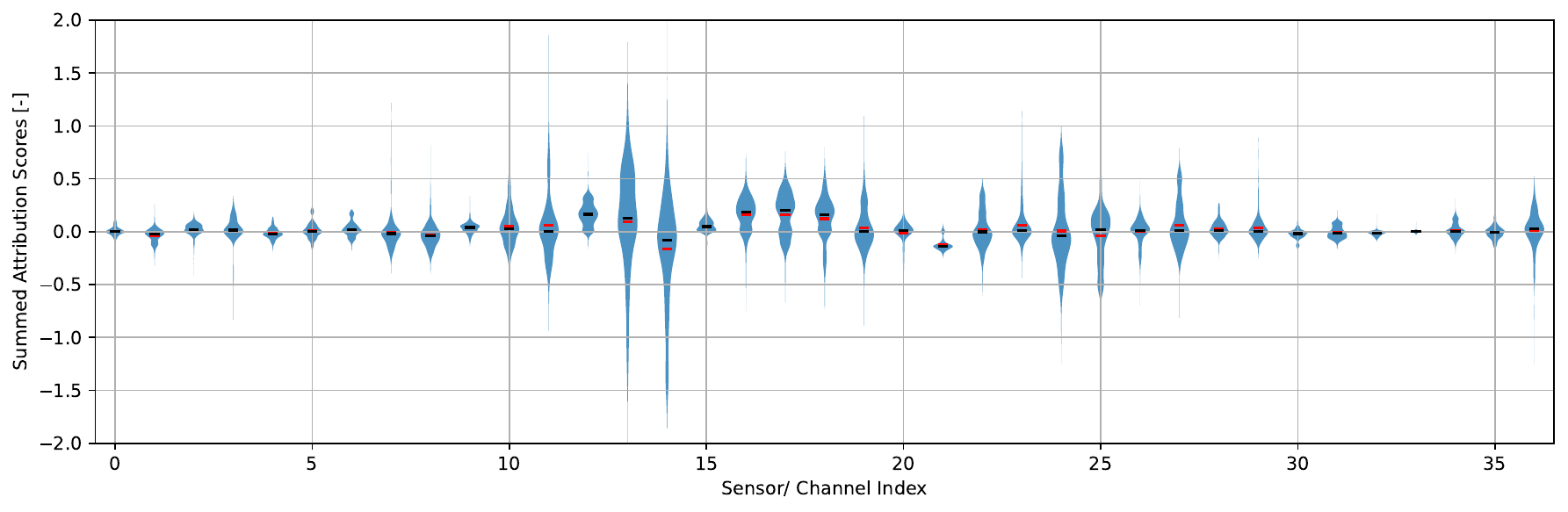}
	\caption{Distribution of summed attribution vectors for the APB and 8° AoA for all correctly classified samples of the validation set (red bar: mean value, black bar: median).}
	\label{fig:channels_stats_bs_apb_8}	
\end{figure}

Figure~\ref{fig:sample_att_bs_apd}~b) displays the attribution map using the APB for the  randomly selected sample~14 of the correctly classified samples of the validation set of split~1. This sample was acquired in TS~1 (see Table~\ref{tab:boundary_conditions}) using an undamaged beam and is visualized in Figure~\ref{fig:signal}. The attribution map covers the full~1.5~s duration of the signal and all~37 operational sensors. The color bar shown in panel~a) is also applicable to panels~b),~d),~e), and~f).
The attribution map is dominated by channels~14-16 which exhibit the largest gradients. Overall, significant IGs are concentrated in two bands between the sensors~9-17 and~23-29 and near zero elsewhere. This is also visible in panel~c) that displays the sum of attributions over the sample duration (compare Equation~\ref{eq:att_vec}), with sensors~14,~15, and~16 highlighted in orange as the largest. This indicates that the predictions of the~0°~AoA CNN are primarily driven by these three adjacent sensors on the leading edge (compare Figure~\ref{fig:sensors_airfoil}), with other channels contributing only weakly, e.g. by delivering an embedding for the information of the dominating channels.
Panels~d)-f) overlay the signals of sensors~14,~15, and~16 with their attribution scores, ordered by absolute sum (sensor~14 highest;~15 lowest). Sensors~15 and~16 exhibit three distinct episodes of mostly positive high-magnitude attributions, while sensor~14 displays four primarily negative episodes. Sensor~9 shows a similar temporal alignment to channels~15 and~16, whereas attributions for other sensors remain diffuse and lack distinctive patterns.
These bursts in IG magnitude suggest that the~0°~AoA CNN learned to respond to local signal morphology likely stemming from cyclic physical phenomena, rather than simple mean levels. However, these subsequences, potentially spanning multiple channels, remain physically difficult to interpret. 
These observations and conclusions hold across other inflow conditions (see Figure~\ref{fig:overview_ts_apb}) and structural states (compare Figure~\ref{fig:overview_dam_apb}).
For~8°~AoA, the attribution maps are similar across inflow conditions and damage states and exhibit minimal temporal variation (see Figures~\ref{fig:overview_ts_apb_8} and~\ref{fig:overview_dam_apb_8}). Aside from slight patterns in TS~5 and~7, the attribution scores remain constant over the sample length, indicating that the CNN, trained on~8°~AoA data, prioritizes static features over temporal dynamics. While leading edge sensors generally exhibit the highest gradients, some inflow conditions result in significant attributions across all sensors between~12 and~28.
The violin plot in Figure~\ref{fig:violin_apb} illustrates the summed attributions per channel (compare Equation~\ref{eq:att_vec}) for all correctly classified validation samples of split~1 with~0°~AoA, with red and black markers denoting the mean and median, respectively. Across all samples, sensors~14-17 near the leading edge exhibit the largest IGs and thus drive the prediction, albeit with the highest variance. Sensors~9-17 and~23-29 generally show IGs of relevant magnitude. The high variability and the bimodal distributions observed for sensors~23-36 likely stem from the diversity of inflow conditions and structural states, and to a lesser extent, from sample-to-sample variance.
For sensors~7-10 and~22-27, gradients vary between positive and negative values suggesting these channels may suppress the predicted class or reflect sub-patterns in specific inflow or structural states. The remaining sensors with mean and median close to zero and a small spread are likely ignored by the model, indicating that they carry little class-discriminative information and may be noisy or redundant for the classification task. These findings also apply to the~8°~AoA attributions based on the APB, whose summed attribution distributions (Figure~\ref{fig:channels_stats_bs_apb_8}) closely resembles the 0° AoA attribution vector distribution, indicating that also the CNN trained on 8° AoA data focuses on the leading edge sensors.
%
\subsection{Attribution maps for the temporal variations baseline} \label{sec_res}
\begin{figure}[h]
    \centering
	\includegraphics[width=\linewidth]{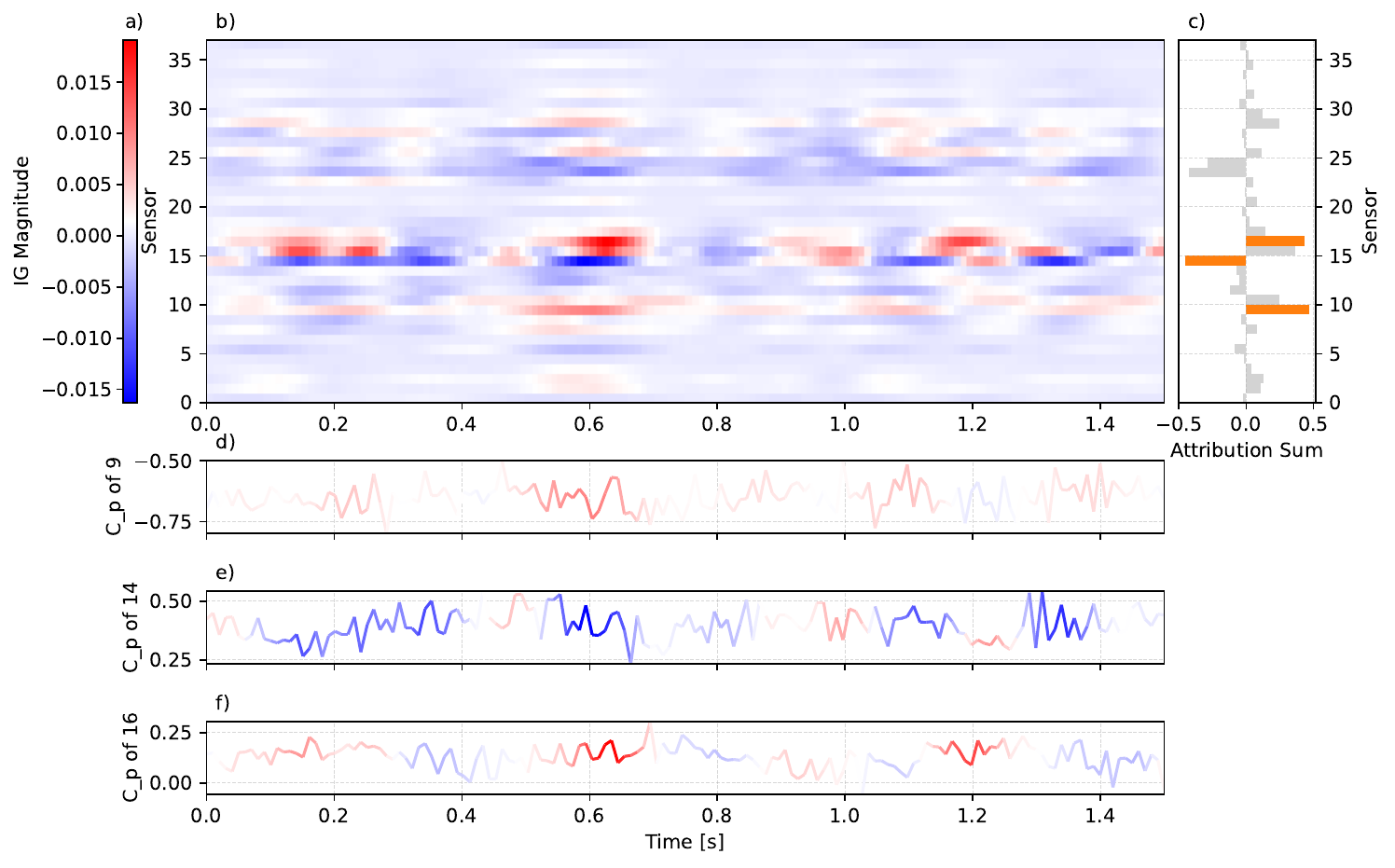}
	\caption{b) Attribution map for sample 14 (split 1, 0° AoA, damage class 0, TS 1) using the TVB. c) Sum of attributions per channel, with the top three highlighted in orange. d)-f) Signals for the three highest attribution channels, colored by attribution score. The colorbar in a) applies
to b) and d)-f).}
	\label{fig:sample_att_bs_TVB}	
\end{figure}

\begin{figure}[h]
    \centering
	\includegraphics[width=\linewidth]{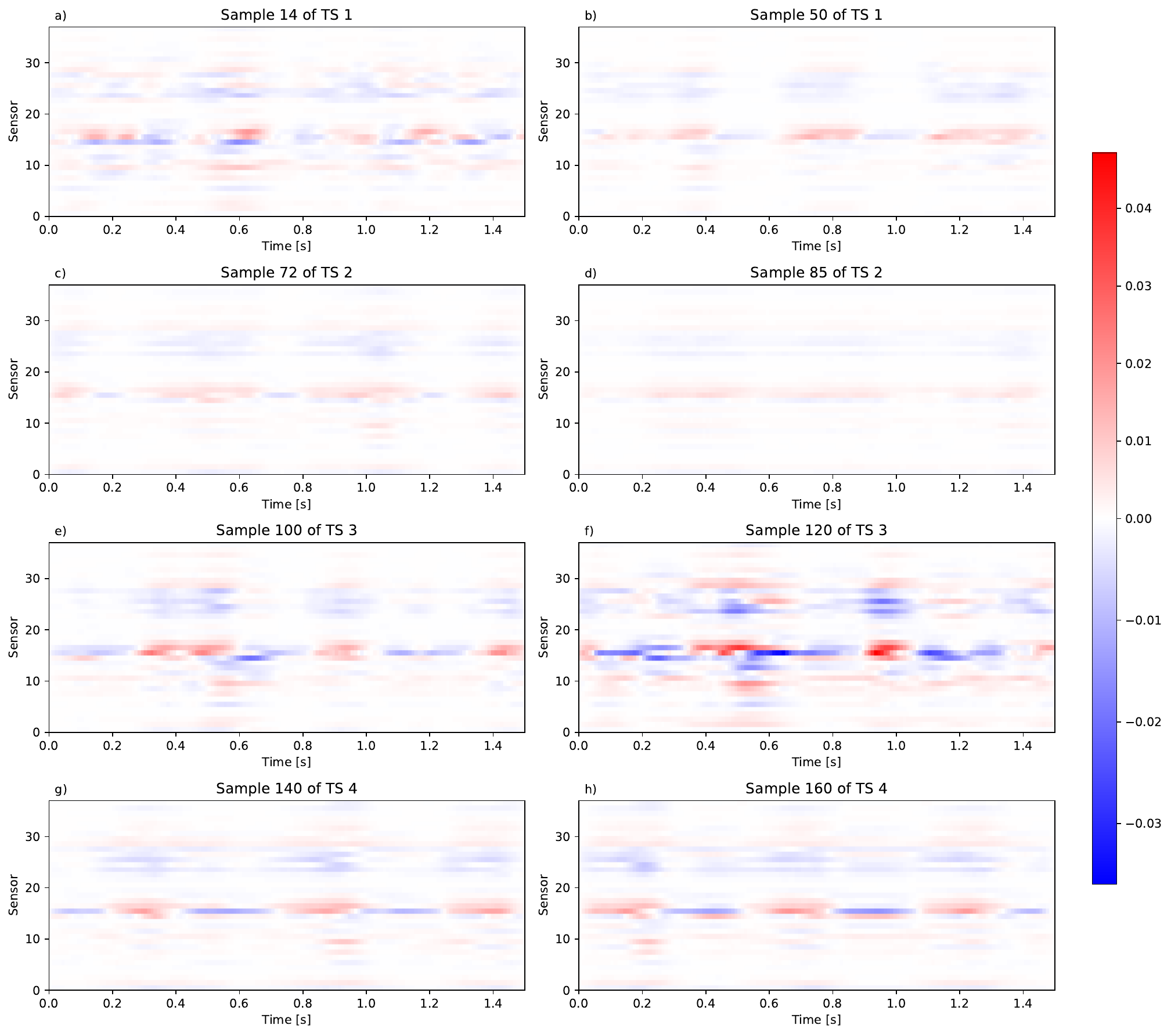}
	\caption{Overview of attribution maps based on the TVB for different inflow conditions for damage class 0 and 0° AoA. Panels a) and b) belong to TS 1 (see Table \ref{tab:boundary_conditions}), c) and d) to TS 2, e) and f) to TS 3 and g) and h) to TS 4.}
	\label{fig:overview_att_bs_tvb}	
\end{figure}

\begin{figure}[h]
    \centering
	\includegraphics[width=0.9\linewidth]{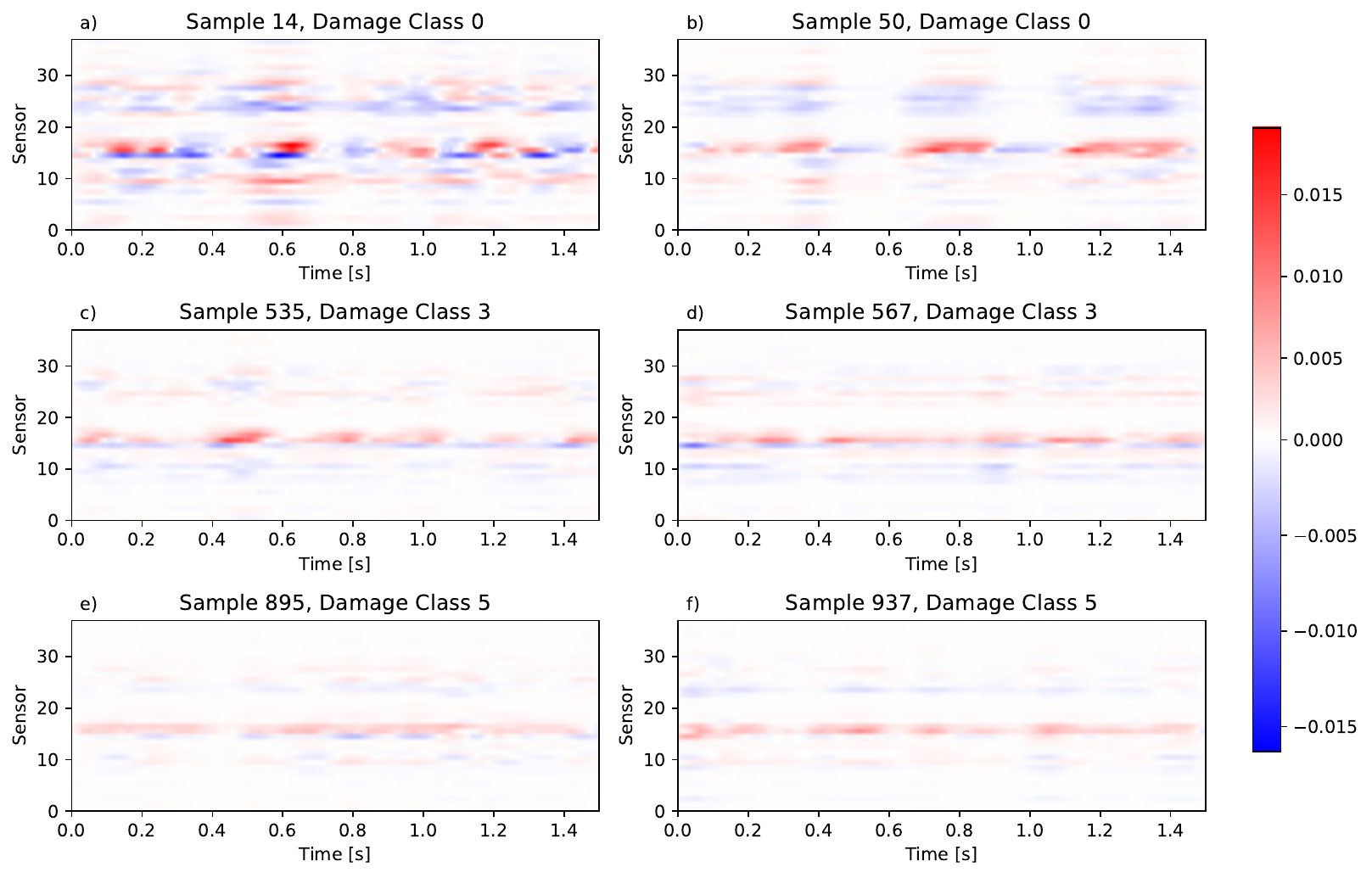}
	\caption{Overview of attribution maps based on the TVB for different structural states for TS 1 and 0° AoA. Panels a) and b) belong to damage class 0 (no damage), c) and d) to damage class 3 (crack length of 25\% of the beam width) and e) and f) to damage class 5 (crack length of 50\% of beam width).}
	\label{fig:overview_att_dam_bs_tvb}	
\end{figure}

\begin{figure}[h]
    \centering
	\includegraphics[width=0.88\linewidth]{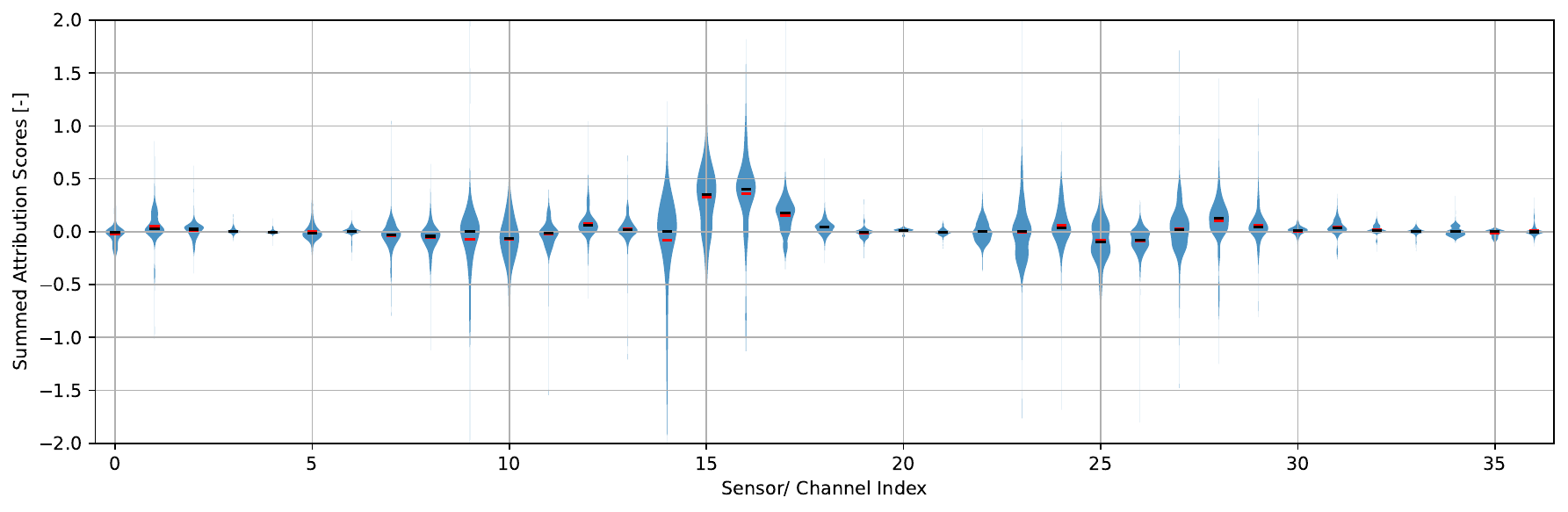}
	\caption{Distribution of summed attribution vectors for the TVB and 0° AoA, for all correctly classified samples of the validation set (red bar: mean value, black bar: median).}
	\label{fig:channels_stats_bs_tvb}	
\end{figure}

\begin{figure}[h]
    \centering
	\includegraphics[width=\linewidth]{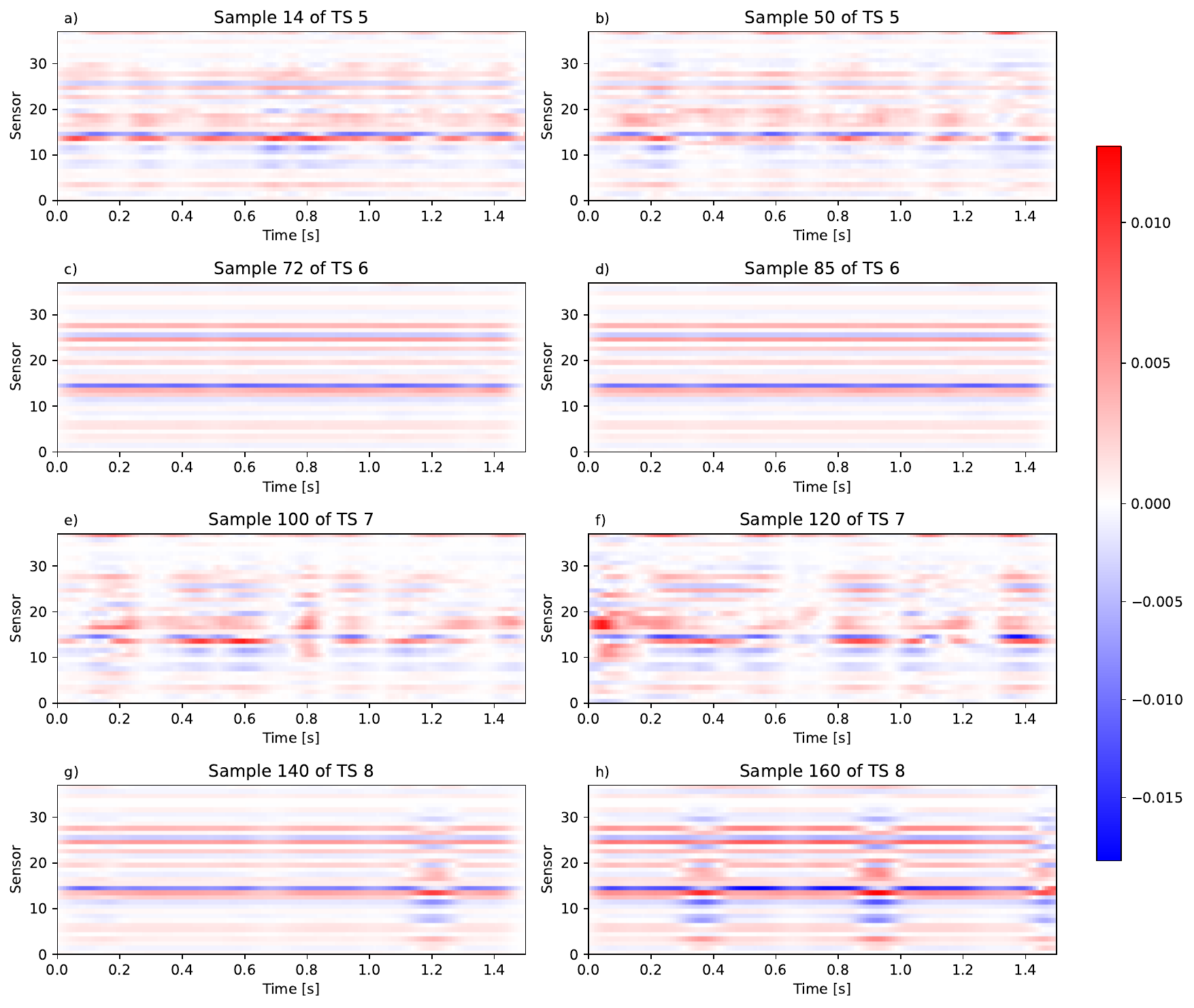}
	\caption{Overview of attribution maps based on the TVB for different inflow conditions for damage class 0 and 8° AoA. Panels a) and b) belong to TS 5 (see Table \ref{tab:boundary_conditions}), c) and d) to TS 6, e) and f) to TS 7 and g) and h) to TS 8.}
	\label{fig:overview_att_bs_tvb_8}	
\end{figure}

\begin{figure}[h]
    \centering
	\includegraphics[width=0.9\linewidth]{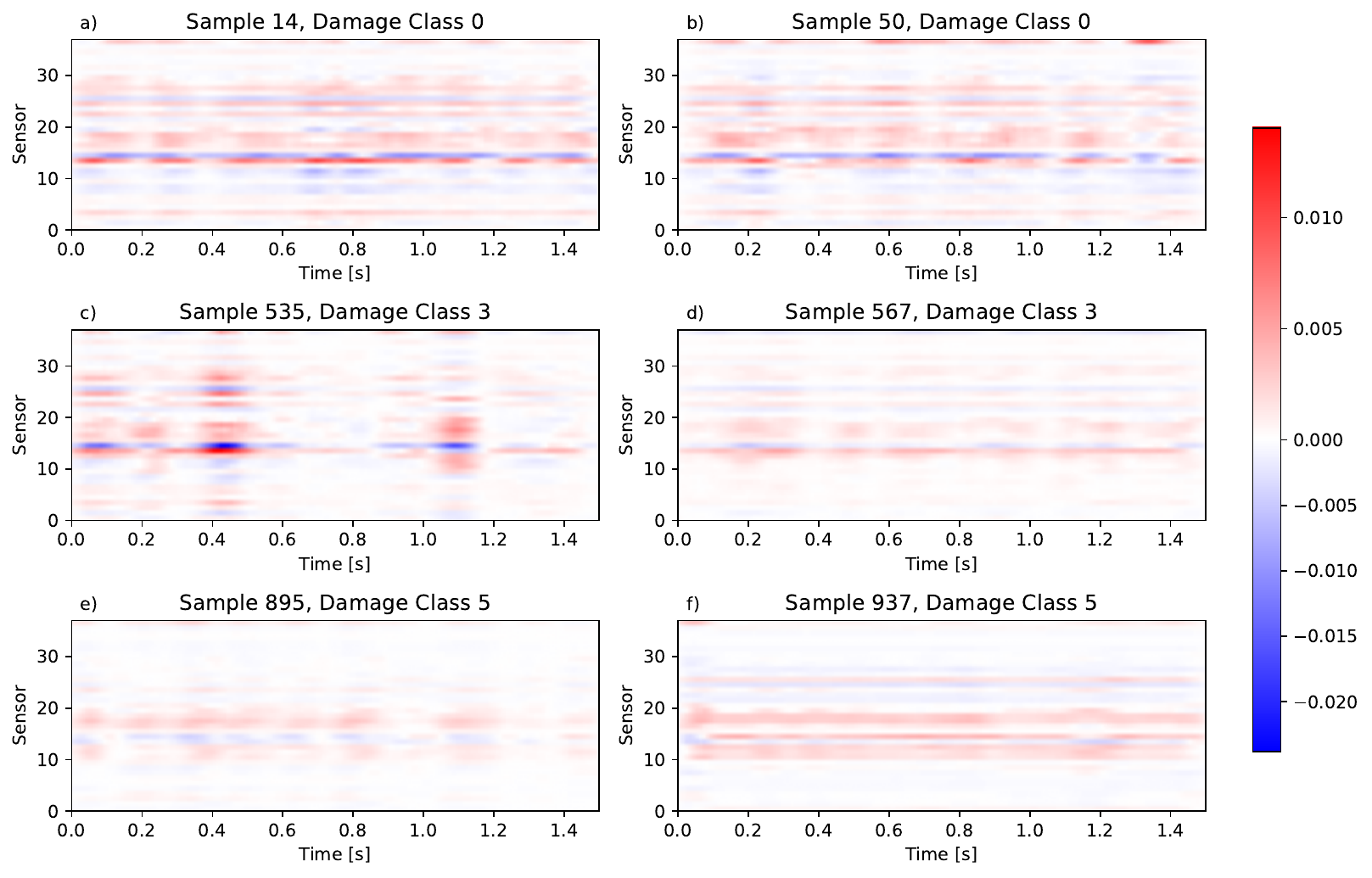}
	\caption{Overview of attribution maps based on the TVB for different structural states for TS 5 and 8° AoA. Panels a) and b) belong to damage class 0 (no damage), c) and d) to damage class 3 (crack length of 25\% of the beam width) and e) and f) to damage class 5 (crack length of 50\% of beam width).}
	\label{fig:overview_att_bs_dam_tvb_8}	
\end{figure}

\begin{figure}[h]
    \centering
	\includegraphics[width=0.88\linewidth]{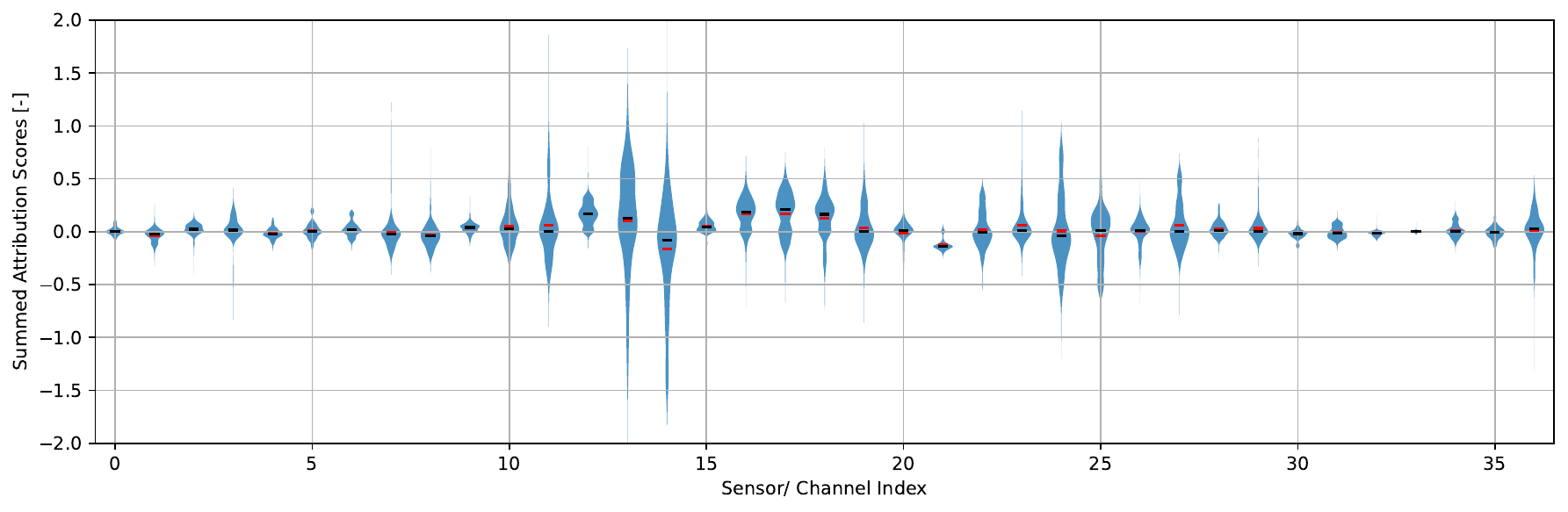}
	\caption{Distribution of summed attribution vectors for the TVB and 8° AoA, for all correctly classified samples of the validation set (red bar: mean value, black bar: median).}
	\label{fig:channels_stats_bs_tvb_8}	
\end{figure}

%
Figure~\ref{fig:sample_att_bs_TVB}~b) displays the attribution map based on the TVB for sample~14 of the correctly classified part of the validation set for split~1 and TS~1.  
Similar to the APB, the TVB attribution maps highlight sensors~9-17 and 23-29 as significant, a finding that remains consistent across different inflow conditions and AoAs, and damage states (compare Figures~\ref{fig:overview_att_bs_tvb},~\ref{fig:overview_att_dam_bs_tvb},~\ref{fig:overview_att_bs_tvb_8} and~\ref{fig:overview_att_bs_dam_tvb_8}).
For~0°~AoA, the subsequences of IG magnitude in channels~14-16 mirror the APB, again suggesting the CNN trained on~0°~AoA relies on recurring multivariate patterns rather than simple mean values, and remain physically difficult to interpret. Conversely, for~8°~AoA (see Figure~\ref{fig:overview_att_bs_tvb_8}), the attribution maps exhibit fewer temporal variations and more constant sections for certain channels. This corroborates the indication from the APB that the~8°~AoA CNN prioritizes static features or general signal characteristics over temporal dynamics.
Evaluating the attribution vectors~$c$ (Equation~\ref{eq:att_vec}) for correctly classified validation samples at 0° AoA via Figure~\ref{fig:channels_stats_bs_tvb} and at~8°~AoA via Figure~\ref{fig:channels_stats_bs_tvb_8} reveals that, analogously to the APB, also for the TVB, both CNNs prioritize leading edge sensors (especially the channels~9-17 and~23-29), while trailing edge sensors (channels~0-8 and~30-36) receive negligible attributions, likely due to minimal contribution to the classification task or their smaller signal offsets.
%
\subsection{Attribution maps for the mean value baseline}
\begin{figure}[h]
    \centering
	\includegraphics[width=\linewidth]{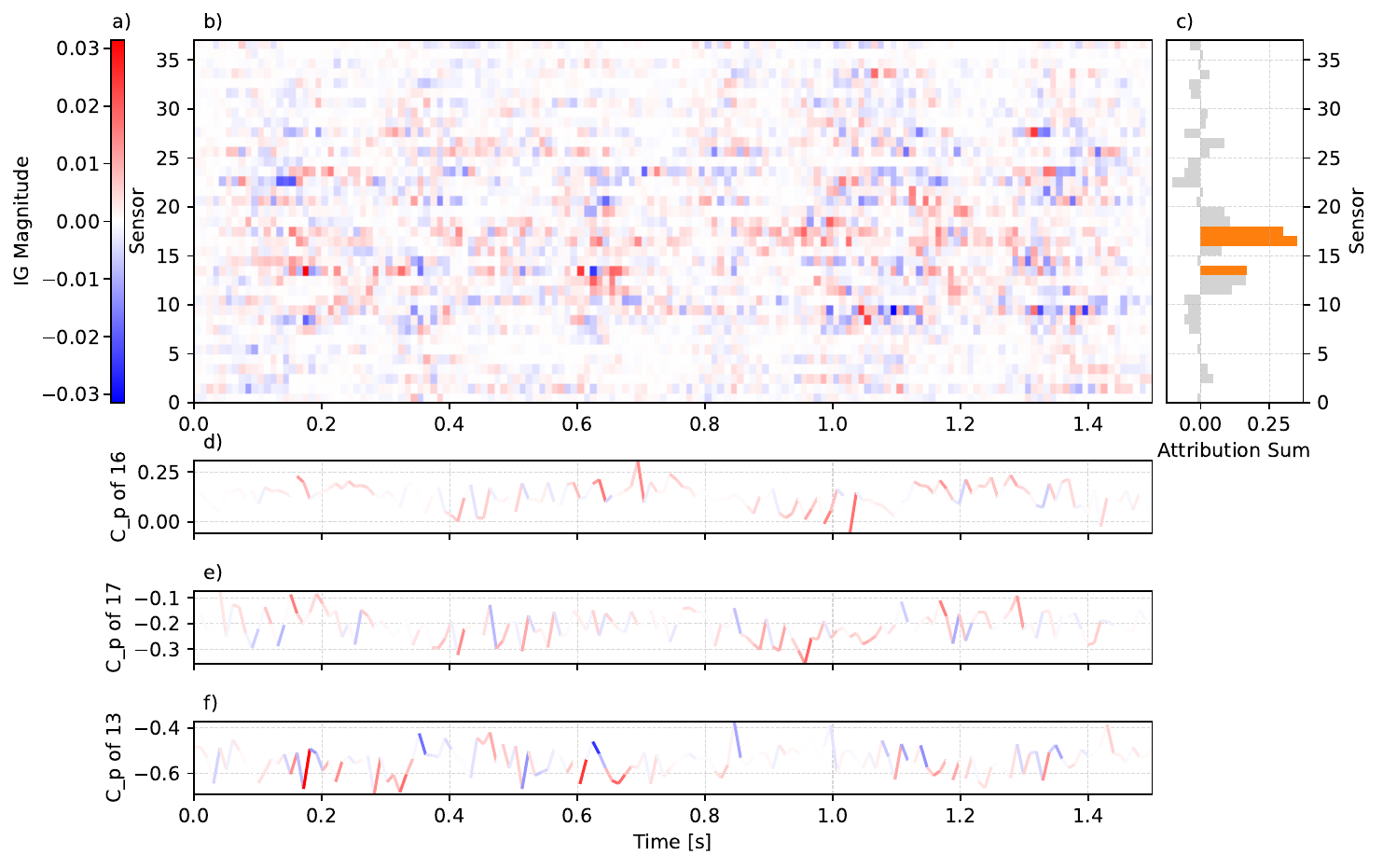}
	\caption{b) Attribution map for sample~14 (split 1, 0° AoA, damage class 0, TS 1) using the MVB. c) Sum of attributions per channel, with the top three channels highlighted in orange. d)-f) Signals for the three highest‑attribution channels, color-coded by attribution score. The colorbar in a) applies to b) and d)-f).}
	\label{fig:sample_att_mvb}	
\end{figure}

\begin{figure}[h]
    \centering
	\includegraphics[width=\linewidth]{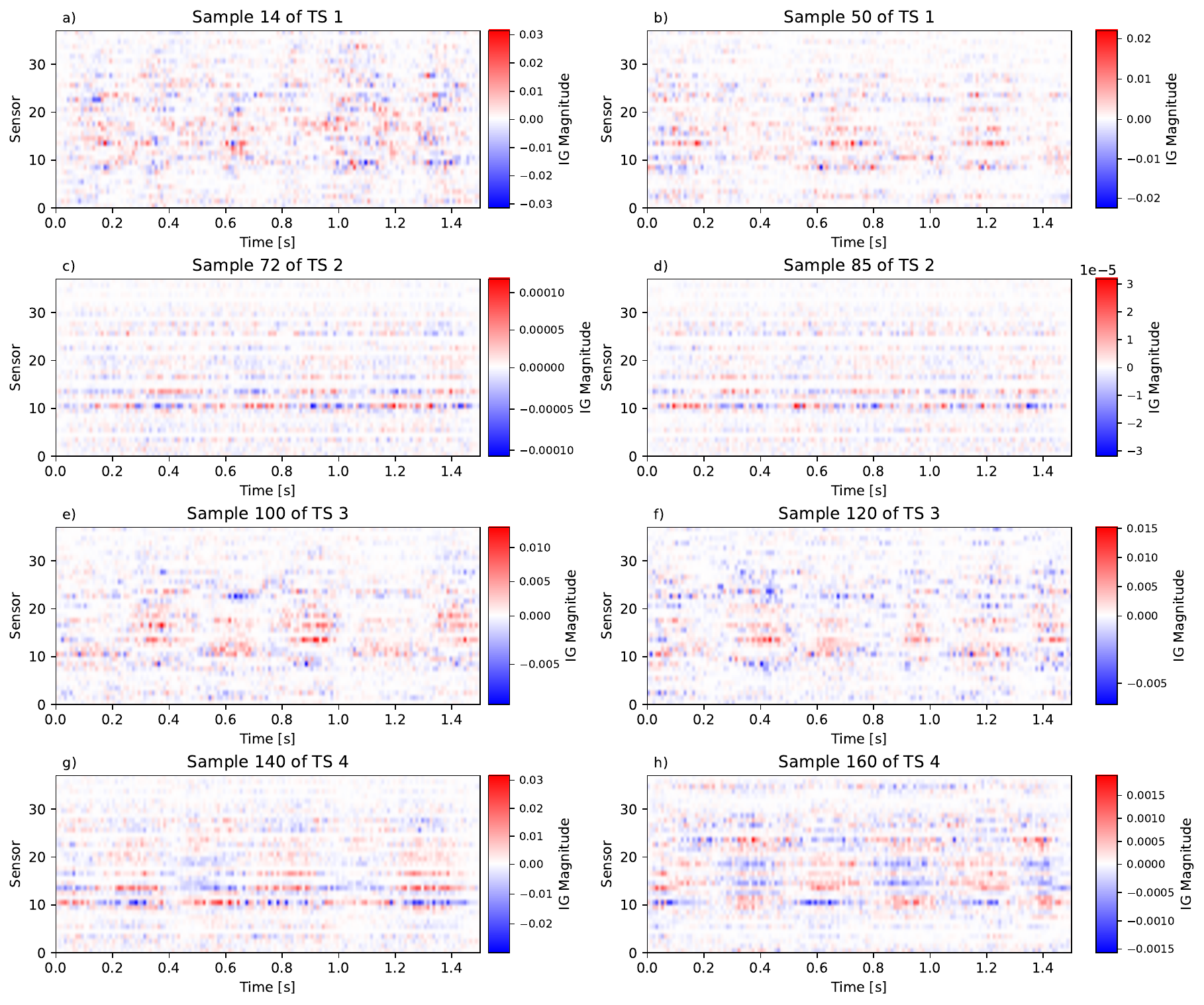}
	\caption{Overview of attribution maps based on the MVB for different inflow conditions for damage class 0 and 0° AoA. Panels a) and b) belong to TS 1 (see Table \ref{tab:boundary_conditions}), c) and d) to TS 2, e) and f) to TS 3 and g) and h) to TS 4.}
	\label{fig:overview_att_mvb}	
\end{figure}

\begin{figure}[h]
    \centering
	\includegraphics[width=0.85\linewidth]{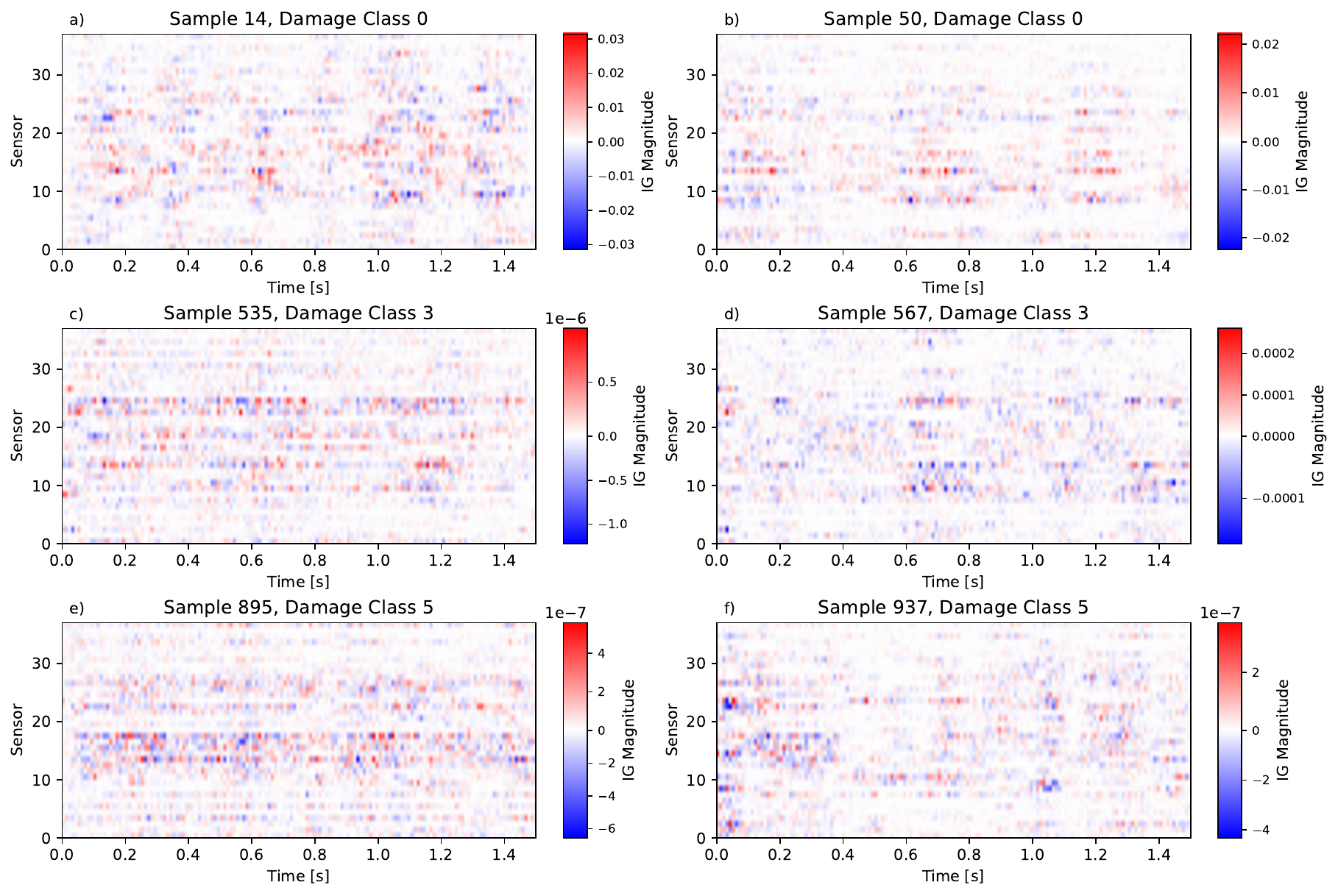}
	\caption{Overview of attribution maps based on the MVB for different structural states for TS 1 and 0° AoA. Panels a) and b) belong to damage class 0 (no damage), c) and d) to damage class 3 (crack length of 25\% of the beam width) and e) and f) to damage class 5 (crack length of 50\% of beam width).}
	\label{fig:overview_att_mvb_damage}	
\end{figure}

\begin{figure}[h]
    \centering
	\includegraphics[width=0.85\linewidth]{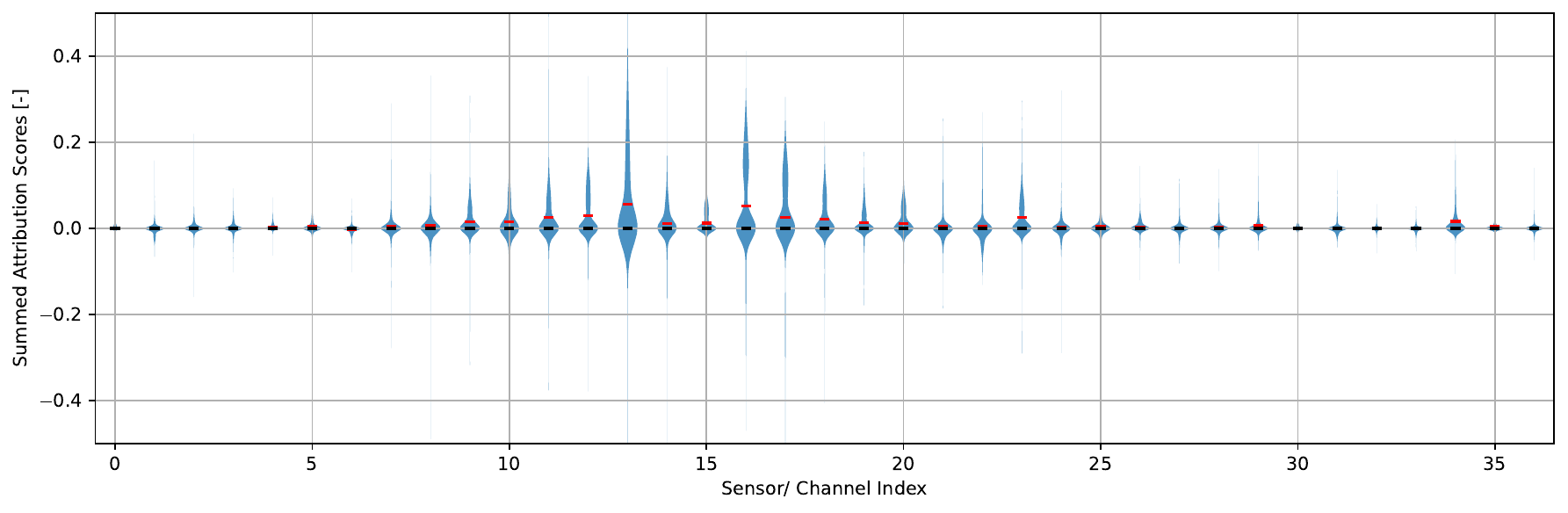}
	\caption{Distribution of summed attribution vectors for the MVB and 0° AoA, for all correctly classified samples of the validation set (red bar: mean value, black bar: median).}
	\label{fig:channels_stats_mvb}	
\end{figure}

\begin{figure}[h]
    \centering
	\includegraphics[width=\linewidth]{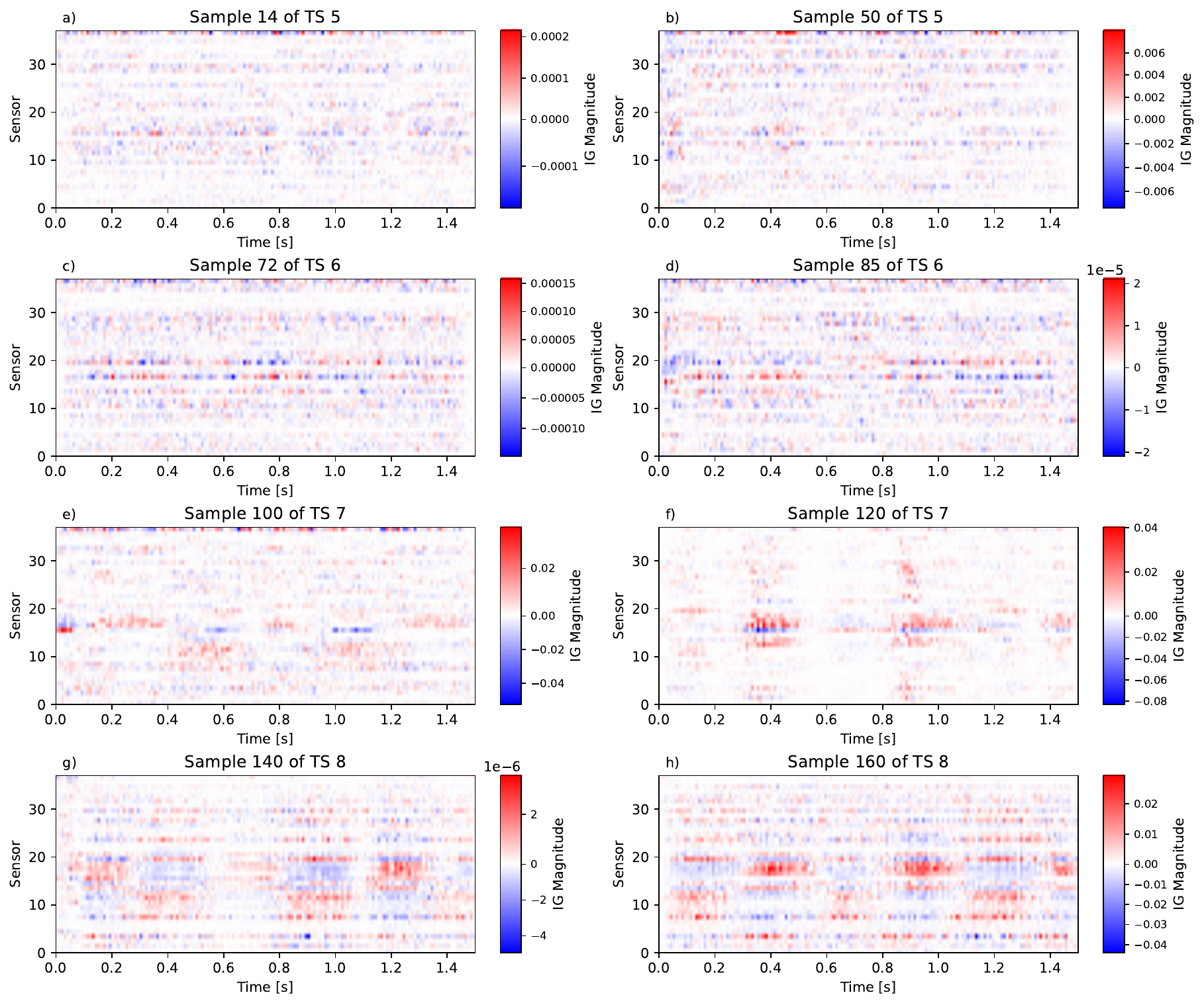}
	\caption{Overview of attribution maps based on the MVB for different inflow conditions for damage class 0 and 8° AoA. Panels a) and b) belong to TS 5 (see Table \ref{tab:boundary_conditions}), c) and d) to TS 6, e) and f) to TS 7 and g) and h) to TS 8.}
	\label{fig:overview_att_mvb_8}	
\end{figure}

\begin{figure}[h]
    \centering
	\includegraphics[width=0.85\linewidth]{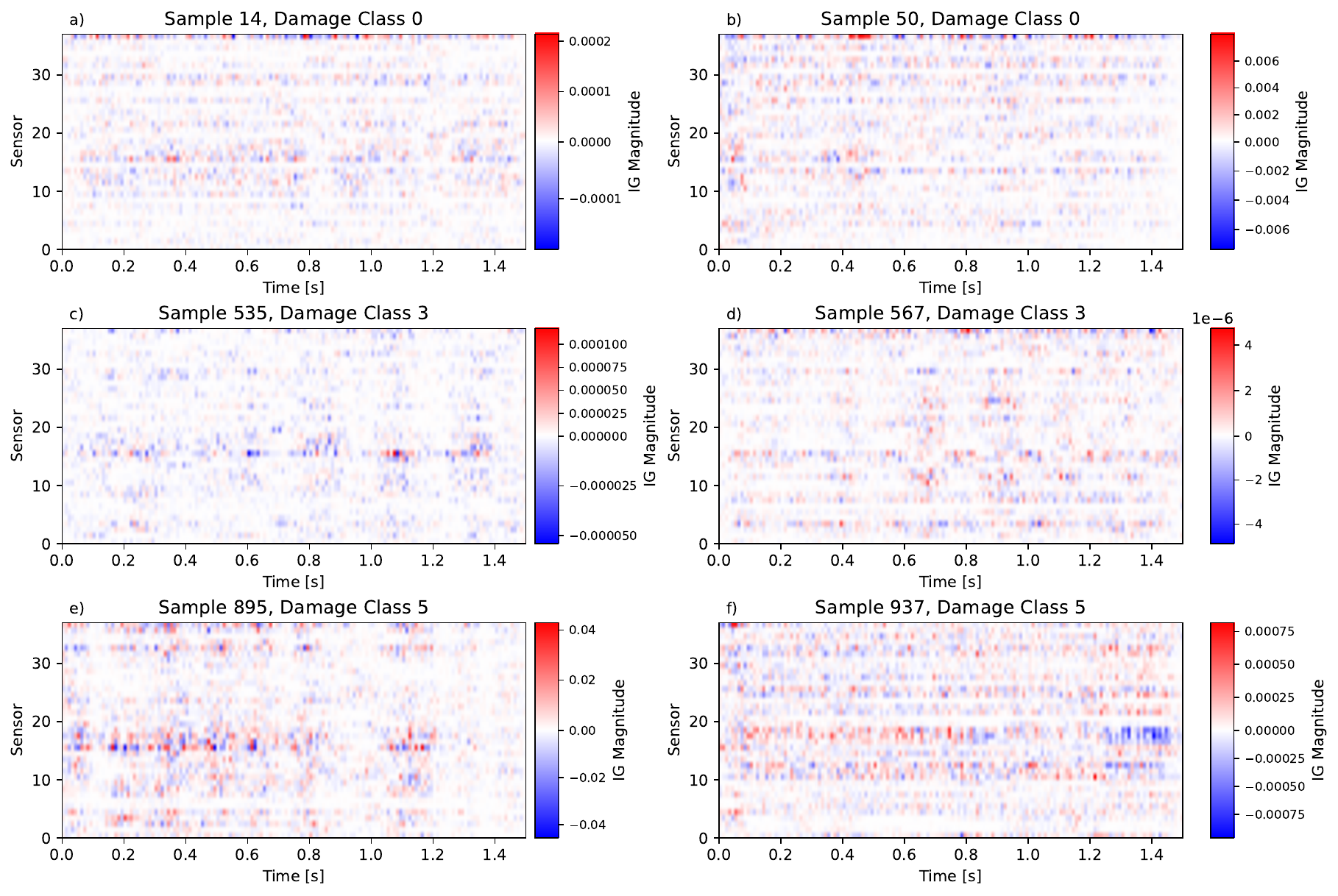}
	\caption{Overview of attribution maps based on the MVB for different structural states for TS 5 and 8° AoA. Panels a) and b) belong to damage class 0 (no damage), c) and d) to damage class 3 (crack length of 25\% of the beam width) and e) and f) to damage class 5 (crack length of 50\% of beam width).}
	\label{fig:overview_att_mvb_damage_8}	
\end{figure}

\begin{figure}[h]
    \centering
	\includegraphics[width=0.85\linewidth]{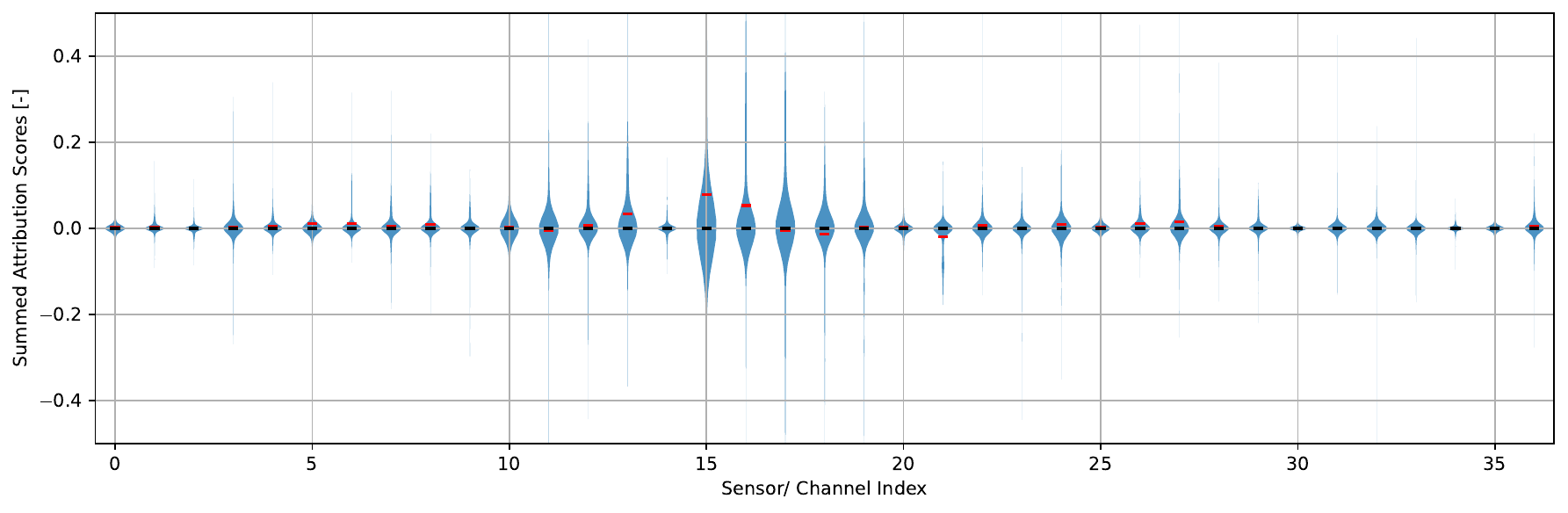}
	\caption{Distribution of summed attribution vectors for the MVB and 8° AoA, for all correctly classified samples of the validation set (red bar: mean value, black bar: median).}
	\label{fig:channels_stats_mvb_8}	
\end{figure}

Figure~\ref{fig:sample_att_mvb} b) illustrates the attribution map using the MVB for the previously introduced sample~14. This map differs significantly from the previous attribution maps, as the IGs are more widely distributed across the entire sample. Furthermore, the concentration of high gradients in channels~14-16 is less pronounced; instead, the higher attribution scores are spread across channels~11-26. Summing the IGs over the sample length shows that channels~13, 16 and 17 exhibit the highest IG sums (see panel~c)), although these sums range approximately from~-0.1 to~0.25, which is lower than those observed for the APB and TVB.
As before, panels~d),~e), and~f) display measurement signals overlaid with attribution scores. The temporal structure of these attributions is scattered, highlighting individual time steps rather than subsequences, unlike the other baselines. Assuming that relevant physical phenomena occur with a frequency smaller than 50 Hz, these should, given our sampling frequency of $f_s=$100 Hz, manifest over time intervals rather than single time steps. Therefore, we attribute these scattered attributions to the baseline choice rather than the underlying physics.
These findings persist across samples recorded under different inflow conditions (see Figure~\ref{fig:overview_att_mvb}) and damage states (see Figure~\ref{fig:overview_att_mvb_damage}). 
The violin plot of the attribution vectors $c$ in Figure~\ref{fig:channels_stats_mvb} reveals notably lower attribution scores compared to the other baselines. While the attribution vectors $c$ of the other baselines span approximately~$[-2; 2]$, the MVB attributions lie mainly in the range of~$[-0.5; 0.5]$. Also for the MVB, the channels~10-17, located around the leading edge, exhibit higher IG sums accompanied by considerable variability. As one can see in Figures~\ref{fig:overview_att_mvb_8}-\ref{fig:channels_stats_mvb_8}, these findings also hold for the~8°~AoA data and CNN.
These findings confirm once more that both CNNs are highly sensitive to signals of the leading edge sensors. 
Overall, the MVB yields for both AoAs substantially lower attribution scores than the APB or TVB. This suggests that the MVB already captures significant class-discriminative information and hints that temporal variations contribute only marginally to the model's predictions.
%
%
\section{Reevaluation Of The CNNs Based On Insights From The Attribution Analysis} \label{sec:class}
%

\subsection{Pre-trained models}
After examining the attribution maps for the different baselines, we now quantitatively assess how the CNNs of~\citep{Franz25}, trained on the original, unaltered recordings, perform when the input is reduced to each baseline and thus lacking a significant signal property of the input samples.
\begin{table}[]
\centering
\caption{Classification accuracy on the test sets for the MVB with CNNs of~\citep{Franz25} that were trained on the unaltered input samples. The results for the APB and TVB are omitted in this table as for these baselines the accuracy for each split is~$\approx$16.67\%.}
\begin{tabular}{@{}ccccc@{}}
\toprule
\multicolumn{5}{c}{MVB}         \\ \midrule
   & Split 1 & Split 2 & Split 3 & Average \\ \midrule
0° AoA & 48.3\% & 53.0\% & 49.9\% & 50.4\%   \\
8° AoA & 61.3\% & 60.2\% & 64.6\% & 62.0\%   \\ \bottomrule
\end{tabular}  \label{tab:bs_pretrained}
\end{table} 
For the APB and TVB, the accuracy of the pre-trained models falls for each split to~$1/6 \approx 16.7\%$, which is indistinguishable from random guessing in a six‑class problem. In contrast, the mean value baseline yields considerably higher scores on all splits (see Table~\ref{tab:bs_pretrained}): about~$50\%$ for the~0°~AoA and roughly~62\% for the~8°~AoA. These results suggest that the previously trained CNNs of~\citep{Franz25} rely primarily on the relative magnitudes across sensors rather than on the temporal dynamics of the signals. This also fits the results obtained by the attribution analysis of the MVB: the scattered and very small attribution scores already hinted that the MVB might contain a significant amount of class-discriminative information.
To verify this hypothesis we will subsequently retrain the CNNs on the baseline data.
\subsection{Retrained models} \label{subsec_baseline_res}
To assess information retention, we retrain the CNN of Section~\ref{subsec:cnn} on the TVB and a regularized MLP on the MVB, collapsed to mean value vectors~$\in\mathbb{R}^{37}$. While an MLP was used to maintain comparability with the CNN, classical learners like random forests~\citep{Breiman.2001} or gradient-boosting machines~\citep{Friedmann.2001} are also viable alternatives.
The complete pipeline, including preprocessing and the MLP architecture, is illustrated in Figure~\ref{fig:mlp} in the appendix. The preprocessing of the measurement data for the MLP is the same as for the CNN approach; only after computing the mean vectors an additional z-scoring is conducted (compare Figure~\ref{fig:mlp}).
The MLP comprises four fully‑connected layers and employs batch normalization and dropout layers to improve generalization and speed up training.
The training protocol is similar to the CNN training and only slightly modified to enhance the stability of the training curve; no hyperparameter optimization was conducted. The full training details are provided in the Appendix~\ref{sec:app_MLP}. 
The classification performance of the CNN for the TVB and of the MLP for the MVB is subsequently reported in Table~\ref{tab:bs_results}. As the classification problem is balanced, we present in Table~\ref{tab:bs_results} only the average accuracy over the six classes for each split of each AoA. 
The APB is omitted in this evaluation, as training a classifier on zero-only samples is not possible. Remarkably, the MLP using only mean values of the MVB yields a high classification accuracy, which is for each split larger than~70\% and on average~86.5\% resp.~83.7\% for~0° and~8°~AoA. The CNN using the TVB samples achieves a considerably lower overall accuracy for each split, with an average of~47.4\% and~34.8\%. 
Although it is likely that with more sophisticated neural networks and training approaches better classification results may be obtained, these results are a strong indicator that especially information encoded by the mean vector of a sample, most likely the inter-channel magnitude relationships, carry significant class-discriminative information.

\begin{table}[h]
\centering
\caption{Classification accuracy on the MVB and TVB test sets of the retrained networks.}
\begin{tabular}{@{}ccccccccc@{}}
\toprule
 & \multicolumn{3}{c}{MVB (MLP)} &  & \multicolumn{3}{c}{TVB (CNN)} &   \\ \midrule
Split   &  1     &   2    &   3    & Average &    1   &    2   &    3   &  Average\\ \midrule
0° AoA  &  74.3\%     &   98.3\%    &   86.8\%    & 86.5\% & 45.3\% & 47.0\%  & 50.0\%  & 47.4 \% \\
8° AoA  &  72.4\%     &   93.8\%    &   85.0\%    & 83.7\% & 33.7\% & 36.5\%  & 34.1\%  & 34.8 \% \\ \bottomrule
\end{tabular} \label{tab:bs_results}
\end{table}

%
%

\section{Discussion}\label{sec:disc}
Across all investigated baselines, the CNNs of~\cite{Franz25} consistently prioritize sensors around the leading edge, particularly channels~14-16, regardless of inflow conditions or AoA. Furthermore, the attribution maps of the APB and TVB reveal that the CNN trained at~0°~AoA,  detects recurring temporal subsequences, whereas for the CNN trained on~8°~AoA data, attribution scores become approximately constant, indicating a reliance on time-independent features of the 8° AoA model. In contrast, the MVB produces scattered IGs without clear temporal patterns and with magnitudes often significantly smaller than those of the APB/TVB for both AoAs. This suggests that the MVB already captures substantial class-discriminative information.
\noindent When evaluated on the TVB, the accuracy of the pre-trained CNNs collapses to~$\approx$16.7\%, while the MVB retains accuracies of~$\approx$50\%~(0° AoA) and~62\%~(8° AoA). Comparing a CNN retrained on TVB with an MLP trained on the MVB mean vectors further demonstrates this disparity: the MLP achieves~86.5\%~(0°~AoA) and~83.7\%~(8° AoA), compared to~47.4\%~(0°~AoA) and~34.8\%~(8°~AoA) for the retrained CNN. 
\noindent Based on these results, we conclude that spatial sensor relationships, here reflected in the channel mean values, constitute the primary source of class-discriminative information. While temporal dynamics also contribute, as evidenced by the TVB accuracies and the superior performance of full-sample classification over mean value vectors, their influence appears secondary. The pronounced performance gap between the MVB and TVB suggests that temporal features are either intrinsically less informative or more difficult to extract, potentially due to signal noise or limitations of the employed neural network architecture. It should be noted that this conclusion may be influenced by the representational capacity of the chosen architecture, which may not optimally capture temporal dependencies.
\noindent Regarding these findings now in the context of the effects of damage on the structure and the APD, the classification accuracy of our CNNs is primarily driven by static, global features and, to a lesser extent, by temporal patterns.
Given the absence of significant vortex shedding in this experimental campaign, these temporal variations are attributed to the oscillation amplitudes of the cantilever beam.
Furthermore, the reliance on static features suggests shifts in the data across structural states, which likely stem from the increasing deformation of the equilibrium position for each state. 
\noindent However, the extent to which the static deformations of the cantilever beam alone account for the observed shifts in the APD remains an open question. Furthermore, the origin of the high-magnitude negative IGs in certain leading edge channels, particularly channel~14, cannot yet be explained from a physical perspective. While the attribution analysis identifies dominant feature patterns, it does not resolve class-specific differences. More structured XAI approaches, such as class-prototype-based methods, may therefore be required to disentangle state-dependent characteristics.
%
\section{Conclusion \& Outlook}\label{sec:concl} 
%

In this paper, we employ IG to enhance the interpretability of the CNN-based structural damage detection approach presented in~\citep{Franz25}. By analyzing the attribution maps from IG based on the APB, TVB, and MVB, we find that the CNNs focus on the signals of sensors located around the leading edge of the airfoil and rely primarily on static, inter-channel magnitude relationships, likely stemming from shifts in the equilibrium position of the structural system, rather than temporal dynamics. While temporal patterns do provide some discriminative information, as evidenced by the performance of the retrained network and the superiority of the CNNs of~\cite{Franz25} over the mean-value MLP, their role is secondary. The exact origin of these static shifts needs to be verified and the high-magnitude negative IGs in leading edge sensors remain unexplained, providing a direction for future experimental design.
Future work should explore more sophisticated baselines, for example in the frequency domain, and more advanced, time-series-specific IG variants~\citep{Jang2025} to extract more targeted features. Additionally, an XAI approach using class-prototypes~\citep{Theissler22} could reveal state-specific features by identifying class representative samples, although high dimensionality and noise remain significant challenges. Finally, developing an unsupervised approach for damage detection and ranking remains a primary objective. \\ \\ \\

\noindent\textbf{Acknowledgements}\\
\indent The authors gratefully acknowledge the computational resources provided through the joint high-performance data analytics (HPDA) project “terrabyte” of the German Aerospace Center~(DLR) and the Leibniz Supercomputing Center (LRZ). AP acknowledges support by dtec.bw~- Digitalization and Technology Research Center of the Bundeswehr (project RISK.twin). dtec.bw is funded by the European Union~-~Next GenerationEU. GD and EC were partially supported by the French-Swiss project MISTERY funded by the French
National Research Agency (ANR PRCI Grant~No.~266157) and the Swiss National Science Foundation (Grant No.~200021L\_212718). \\ \\

\noindent\textbf{Use of AI tools declaration} \\
\indent The authors declare they have used Artificial Intelligence (AI) tools in the creation of this article: In some paragraphs throughout the article, AI tools have been used to rephrase and improve sentences grammatically and language-wise. We emphasize that no content has been generated by AI tools.

\newpage 
\bibliographystyle{unsrtnat}
\bibliography{references} 

\newpage
\section{Appendix}
\subsection{Structural states \& damage classes}

\begin{table}[h]
\centering 
\caption{Mapping of structural states to damage classes from~\cite{Franz25}.}
\begin{tabular}{@{}lllllll@{}}
\toprule
 Damage Class & 0 & 1 & 2 & 3 & 4 & 5 \\ \midrule
Crack length in \\
$[$\% of beam-width$]$ & 0\% & 0\% and added mass & 12.5\% & 25\% & 37.5\% & 50\%\\
\bottomrule
\end{tabular}
\label{tab:damage_classes}
\end{table}

\subsection{Spectra of a pressure signal recorded near the trailing edge}

\clearpage

\subsection{Integration Path in the Sample Space for Integrated Gradients}

\clearpage
\subsection{Plots for the ambient pressure baseline} \label{app:apb_plots}
\subsubsection{Attribution maps for 0° AoA and varying inflow conditions}
\clearpage

\subsubsection{Attribution maps for 0° AoA and varying structural states}
\clearpage

\subsubsection{Attribution maps for 8° AoA and Varying Inflow Conditions}
\clearpage

\subsubsection{Attribution maps for 8° AoA and Varying Structural States}

\subsubsection{Distribution of summed attribution vectors for 8° AoA}

\clearpage
\subsection{Plots for the temporal variations baseline} \label{app:tvb_plots}

\subsubsection{Attribution maps for 0° AoA and varying inflow conditions}
\clearpage

\subsubsection{Attribution maps for 0° AoA and varying structural states}

\subsubsection{Distribution of summed attribution vectors for 0° AoA}
\clearpage

\subsubsection{Attribution Maps for 8° AoA and varying inflow conditions}
\clearpage

\subsubsection{Attribution Maps for 8° AoA and varying structural states}
\subsubsection{Distribution of summed attribution vectors for 8° AoA}
\clearpage
\subsection{Plots for the mean value baseline} \label{app:mvb_plots}

\subsubsection{Attribution maps for 0° AoA and varying inflow conditions}
\clearpage

\subsubsection{Attribution maps for 0° AoA and varying structural states}

\subsubsection{Distribution of summed attribution vectors for 0° AoA}
\clearpage 

\subsubsection{Attribution maps for 8° AoA and varying inflow conditions}
\clearpage

\subsubsection{Attribution maps for 8° AoA and varying structural states}

\subsubsection{Distribution of Summed Attribution Vectors for 8° AoA}
\clearpage

\subsection{Training details of the MLP} \label{sec:app_MLP}
\noindent Analogously to the CNN approach, the raw APM time series are first segmented into 1.5 s windows and normalized on a per‑sample basis. For each window we compute the mean vector across time and subsequently the resulting vectors for the training, validation, and test set (splits identical to those used for the CNN) are z‑normalized using the training set mean and standard deviation. 
The MLP comprises four fully‑connected layers (widths: 128, 128, 64 and 6) that operate on the normalized mean vectors to produce the final class predictions. To improve generalization and speed up training, each hidden layer is followed by a batch normalization layer, which stabilizes the distribution of activations, and the input and the first two hidden layers are followed by dropout layers, which randomly deactivate a fraction of neurons. The hyperparameters were selected to obtain a stable training curve; no hyperparameter optimization was performed.  
\begin{table}[h]
\centering
\caption{Training hyperparameters for the MLP.}
\label{tab:training-params}
\begin{tabular}{ll}
\toprule
\textbf{Parameter} & \textbf{Value} \\ \hline
Optimizer & AdamW \\
Learning rate & \(1\times10^{-3}\) \\
Weight decay & \(1\times10^{-5}\) \\
Loss function & categorical crossentropy, \\
              & with label smoothing (0.05) \\
Batch size & 32 \\
Maximum epochs & 150 \\
\hline
Callbacks on validation loss: & \\
ModelCheckpoint - save best only& \\
ReduceLROnPlateau - factor=0.5, patience=10, min lr=1e-5 &\\
EarlyStopping - patience=12, restore best weights & \\
\bottomrule
\end{tabular}
\end{table}

\begin{figure}[h]
\centering
\begin{tikzpicture}[node distance=1cm]

\tikzstyle{process_step} = [rectangle, rounded corners=3pt,
minimum width=2cm, 
minimum height=0.8cm,
text centered, 
draw=black]

\tikzstyle{node} = [rectangle, rounded corners=3pt, 
minimum width=1 cm, 
minimum height=0.8cm,
text centered, 
draw=black]

\tikzstyle{layer} = [rectangle, rounded corners=3pt,
minimum width=0.3 cm, 
minimum height=3.0cm,
text centered, 
draw=black,
fill=orange!30]

\tikzstyle{bn_relu} = [rectangle, rounded corners=3pt,
minimum width=0.3 cm, 
minimum height=2.0cm,
text centered, 
draw=black,
fill=purple!25]

\tikzstyle{label} = [rectangle, rounded corners=3pt,
minimum width=0.3 cm, 
minimum height=2.0cm,
text centered, 
draw=black]

\tikzstyle{dropout} = [rectangle, rounded corners=3pt,
minimum width=0.3 cm, 
minimum height=3.0cm,
text centered, 
draw=black,
fill = brown!25]


\node(Input)[data, process_step]{Time Series Data};
\node(SaScaling)[process, process_step, below = 0.5 cm of Input, align=center]{Windowing \& \\ Samplewise scaling};
\node(Mean)[process, process_step, below = 0.5 cm of SaScaling, align=center]{Mean Vectors};
\node(Norm)[process, process_step, below = 0.5 cm of Mean, align=center]{z - Normalization};
\node(DO1)[dropout, above = 0.2 cm of Norm, xshift = 3 cm]{\rotatebox{90}{Dropout, 0.2}};
\node(Conv1)   [layer, right = 0.5 cm of DO1]{\rotatebox{90}{Dense, 128}};
\node(BN_Relu1)[bn_relu, right = -0.01 cm of Conv1]{\rotatebox{90}{BN + ReLU}};
\node(DO2)     [dropout, right = 0.5 cm of BN_Relu1]{\rotatebox{90}{Dropout, 0.4}};
\node(Conv2)   [layer, right = 0.5 cm of DO2]{\rotatebox{90}{Dense, 128}};
\node(BN_Relu2)[bn_relu, right = -0.01 cm of Conv2]{\rotatebox{90}{BN + ReLU}};
\node(DO3)     [dropout, right = 0.5 cm of BN_Relu2]{\rotatebox{90}{Dropout, 0.4}};
\node(Conv3)   [layer, right = 0.5 cm of DO3]{\rotatebox{90}{Dense, 64}};
\node(BN_Relu3)[bn_relu, right = -0.01 cm of Conv3]{\rotatebox{90}{BN + ReLU}};
\node(Gap)     [layer, right = 0.5 of BN_Relu3]{\rotatebox{90}{Dense, 6}};
\node(output)  [output, label, right = 0.5 cm of Gap]{\rotatebox{90}{Label}};

\draw [arrow] (Input) -- (SaScaling);
\draw [arrow] (SaScaling) -- (Mean);
\draw [arrow] (Mean) -- (Norm);
\draw [arrow] (Norm.east) -- ++(0.7,0) |- (DO1.west);
\draw [arrow] (DO1) -- (Conv1);
\draw [arrow] (BN_Relu1) -- (DO2);
\draw [arrow] (DO2) -- (Conv2);
\draw [arrow] (BN_Relu2) -- (DO3);
\draw [arrow] (DO3) -- (Conv3);
\draw [arrow] (BN_Relu3) -- (Gap);
\draw [arrow] (Gap) -- (output);

\node[draw=black!40, dashed, rounded corners,
      fit=(Input)(Norm),
      inner sep=0.25cm] (prepro) {};

\node[above=2pt of prepro] {\textbf{Preprocessing}};

\node[draw=black!40, dashed, rounded corners,
      fit=(DO1)(output),
      inner sep=0.25cm] (classifier) {};

\node[above=2pt of classifier] {\textbf{Classifier}};
\end{tikzpicture}
\caption{Preprocessing and structure for the mean vector MLP.} \label{fig:mlp}
\end{figure}
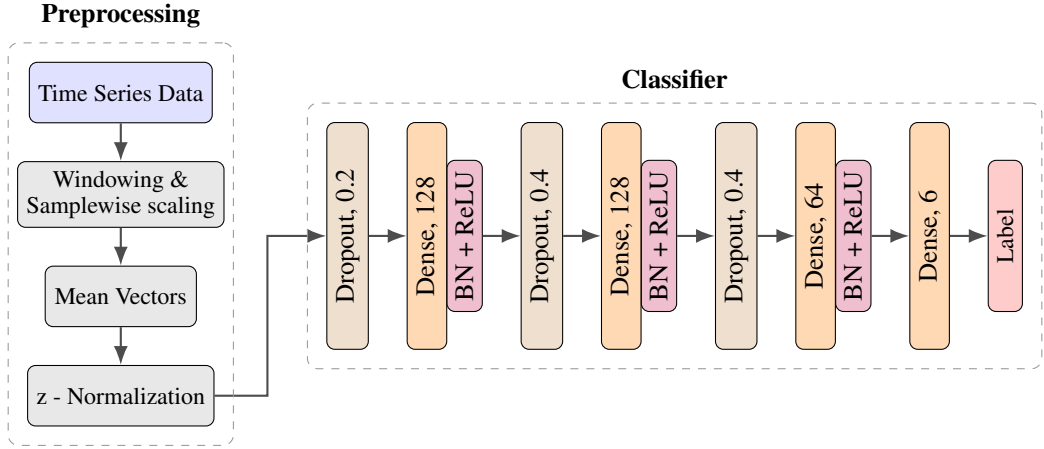  

\end{document}

%% file: airfoil_heaving_twisting.pdf_tex
\begingroup%
  \makeatletter%
  \providecommand\color[2][]{%
    \errmessage{(Inkscape) Color is used for the text in Inkscape, but the package 'color.sty' is not loaded}%
    \renewcommand\color[2][]{}%
  }%
  \providecommand\transparent[1]{%
    \errmessage{(Inkscape) Transparency is used (non-zero) for the text in Inkscape, but the package 'transparent.sty' is not loaded}%
    \renewcommand\transparent[1]{}%
  }%
  \providecommand\rotatebox[2]{#2}%
  \newcommand*\fsize{\dimexpr\f@size pt\relax}%
  \newcommand*\lineheight[1]{\fontsize{\fsize}{#1\fsize}\selectfont}%
  \ifx\svgwidth\undefined%
    \setlength{\unitlength}{1289.71325972bp}%
    \ifx\svgscale\undefined%
      \relax%
    \else%
      \setlength{\unitlength}{\unitlength * \real{\svgscale}}%
    \fi%
  \else%
    \setlength{\unitlength}{\svgwidth}%
  \fi%
  \global\let\svgwidth\undefined%
  \global\let\svgscale\undefined%
  \makeatother%
  \begin{picture}(1,0.42251012)%
    \lineheight{1}%
    \setlength\tabcolsep{0pt}%
    \put(0,0){\includegraphics[width=\unitlength,page=1]{airfoil_heaving_twisting.pdf}}%
    \put(0.38181739,0.2806924){\color[rgb]{0,0,0}\makebox(0,0)[lt]{\lineheight{0}\smash{\begin{tabular}[t]{l}$K_y$\end{tabular}}}}%
    \put(0.46773292,0.08224527){\color[rgb]{0,0,0}\makebox(0,0)[lt]{\lineheight{0}\smash{\begin{tabular}[t]{l}$K_\phi, D_\phi$\end{tabular}}}}%
    \put(0.52476597,0.28129403){\color[rgb]{0,0,0}\makebox(0,0)[lt]{\lineheight{0}\smash{\begin{tabular}[t]{l}$D_y$\end{tabular}}}}%
    \put(0.53859409,0.14843052){\color[rgb]{0,0,0}\makebox(0,0)[lt]{\lineheight{0}\smash{\begin{tabular}[t]{l}$M$\end{tabular}}}}%
    \put(0,0){\includegraphics[width=\unitlength,page=2]{airfoil_heaving_twisting.pdf}}%
    \put(0.07259872,0.15064303){\color[rgb]{0,0,0}\makebox(0,0)[lt]{\lineheight{0}\smash{\begin{tabular}[t]{l}$V$\end{tabular}}}}%
    \put(0.09090932,0.01770456){\color[rgb]{0,0,0}\makebox(0,0)[lt]{\lineheight{0}\smash{\begin{tabular}[t]{l}$V_{rel}$\end{tabular}}}}%
    \put(0.31124888,0.04379568){\color[rgb]{0,0,0}\makebox(0,0)[lt]{\lineheight{0}\smash{\begin{tabular}[t]{l}$-\dot{y}$\end{tabular}}}}%
    \put(0.18948518,0.11051782){\color[rgb]{0,0,0}\makebox(0,0)[lt]{\lineheight{0}\smash{\begin{tabular}[t]{l}$\alpha$\end{tabular}}}}%
    \put(0.32346501,0.23217362){\color[rgb]{0,0,0}\makebox(0,0)[lt]{\lineheight{0}\smash{\begin{tabular}[t]{l}$F_L$\end{tabular}}}}%
    \put(0.49927199,0.19214968){\color[rgb]{0,0,0}\makebox(0,0)[lt]{\lineheight{0}\smash{\begin{tabular}[t]{l}$F_D$\end{tabular}}}}%
    \put(0.72945927,0.25136117){\color[rgb]{0,0,0}\makebox(0,0)[lt]{\lineheight{0}\smash{\begin{tabular}[t]{l}$y$, $\ddot{y}$\end{tabular}}}}%
    \put(0,0){\includegraphics[width=\unitlength,page=3]{airfoil_heaving_twisting.pdf}}%
    \put(0.54363405,0.00866447){\color[rgb]{0,0,0}\makebox(0,0)[lt]{\lineheight{0}\smash{\begin{tabular}[t]{l}$c$\end{tabular}}}}%
    \put(0,0){\includegraphics[width=\unitlength,page=4]{airfoil_heaving_twisting.pdf}}%
    \put(0.74114293,0.29726417){\color[rgb]{0,0,0}\makebox(0,0)[lt]{\lineheight{0}\smash{\begin{tabular}[t]{l}$\phi$, $\ddot{\phi}$\end{tabular}}}}%
  \end{picture}%
\endgroup%

%% file: integration_paths.pdf_tex
\begingroup%
  \makeatletter%
  \providecommand\color[2][]{%
    \errmessage{(Inkscape) Color is used for the text in Inkscape, but the package 'color.sty' is not loaded}%
    \renewcommand\color[2][]{}%
  }%
  \providecommand\transparent[1]{%
    \errmessage{(Inkscape) Transparency is used (non-zero) for the text in Inkscape, but the package 'transparent.sty' is not loaded}%
    \renewcommand\transparent[1]{}%
  }%
  \providecommand\rotatebox[2]{#2}%
  \newcommand*\fsize{\dimexpr\f@size pt\relax}%
  \newcommand*\lineheight[1]{\fontsize{\fsize}{#1\fsize}\selectfont}%
  \ifx\svgwidth\undefined%
    \setlength{\unitlength}{298.55320728bp}%
    \ifx\svgscale\undefined%
      \relax%
    \else%
      \setlength{\unitlength}{\unitlength * \real{\svgscale}}%
    \fi%
  \else%
    \setlength{\unitlength}{\svgwidth}%
  \fi%
  \global\let\svgwidth\undefined%
  \global\let\svgscale\undefined%
  \makeatother%
  \begin{picture}(1,0.3736072)%
    \lineheight{1}%
    \setlength\tabcolsep{0pt}%
    \put(0,0){\includegraphics[width=\unitlength,page=1]{integration_paths.pdf}}%
    \put(0.10040732,0.34076198){\color[rgb]{0,0,0}\makebox(0,0)[t]{\lineheight{1.25}\smash{\begin{tabular}[t]{c}$x'_1,x'_2$\end{tabular}}}}%
    \put(0.907718,0.00941195){\color[rgb]{0,0,0}\makebox(0,0)[t]{\lineheight{1.25}\smash{\begin{tabular}[t]{c}$x_1,x_2$\end{tabular}}}}%
    \put(0.52737996,0.33512839){\color[rgb]{0,0,0}\makebox(0,0)[t]{\lineheight{1.25}\smash{\begin{tabular}[t]{c}$P_1$\end{tabular}}}}%
    \put(0.73087028,0.12656303){\color[rgb]{0,0,0}\makebox(0,0)[t]{\lineheight{1.25}\smash{\begin{tabular}[t]{c}$P_2$\end{tabular}}}}%
    \put(0.50927625,0.23284751){\color[rgb]{0,0,0}\makebox(0,0)[t]{\lineheight{1.25}\smash{\begin{tabular}[t]{c}$P_3$\end{tabular}}}}%
  \end{picture}%
\endgroup%